\documentclass{mn2e}
\usepackage{graphicx}
\usepackage{mncite}

\title[The CVs AC~Cnc \& V363~Aur]
{The masses of the cataclysmic variables AC~Cancri and V363~Aurigae}

\author[T. D. Thoroughgood et al.]
{T. D. Thoroughgood,$^{1}$\thanks{E-mail: Tim.Thoroughgood@shef.ac.uk} 
V. S. Dhillon,$^{1}$ 
C. A. Watson,$^{1}$
D. A. H. Buckley,$^{2}$
\newauthor
D. Steeghs,$^{3}$ 
M. J. Stevenson,$^{1}$ 
\\ \\
$^{1}$Department of Physics and Astronomy, University of Sheffield, 
Sheffield, S3 7RH, UK \\
$^{2}$South African Astronomical Observatory, PO Box 9, Observatory 7935, Cape 
Town, South Africa\\
$^{3}$Harvard-Smithsonian Center for Astrophysics, 60 Garden Street, MS-67, 
Cambridge, MA 02138, USA \\
}

\date{\center{\Large Accepted for publication in the Monthly 
Notices of the Royal Astronomical Society \\ 
\vspace{.5cm} \today}}

\begin{document}
\maketitle

\begin{abstract}
We present time--resolved spectroscopy and photometry of the 
double--lined eclipsing cataclysmic variables AC~Cnc and V363~Aur (= Lanning 10). 
There is evidence of irradiation on the inner hemisphere of the secondary star 
in both systems, which we correct for using a model that reproduces the 
observations remarkably well. 
We find the radial velocity of the secondary star in AC~Cnc to be $K_R$ = 176 
$\pm$ 3 km\,s$^{-1}$ and its rotational velocity to be $v \sin i$ = 135 
$\pm$ 3 km\,s$^{-1}$. From these parameters we obtain masses of 
$M_1$ = 0.76 $\pm$ 0.03 $M_\odot$ for the white dwarf primary and 
$M_2$ = 0.77 $\pm$ 0.05 $M_\odot$ for the K2 $\pm$ 1V secondary star, giving a 
mass ratio of $q$ = 1.02 $\pm$ 0.04. 
We measure the radial and rotational velocites of the G7 $\pm$ 2V secondary 
star in V363~Aur to be $K_R$ = 168 $\pm$ 5 km\,s$^{-1}$ and 
$v \sin i$ = 143 $\pm$ 5 km\,s$^{-1}$ respectively. The component 
masses of V363~Aur are $M_1$ = 0.90 $\pm$ 0.06 $M_\odot$ and 
$M_2$ = 1.06 $\pm$ 0.11 $M_\odot$, giving a mass ratio of $q$ = 1.17 $\pm$ 0.07. 
The mass ratios for AC~Cnc and V363~Aur fall within 
the theoretical limits for dynamically and thermally stable mass transfer.
Both systems are similar to the SW~Sex stars, exhibiting 
single--peaked emission lines with transient absorption features, 
high--velocity S--wave components and phase--offsets in their radial 
velocity curves. 
The Balmer lines in V363~Aur show a rapid increase in flux around phase 0 
followed by a rapid decrease, 
which we attribute to the eclipse of an optically thick 
region at the centre of the disc. This model could also 
account for the behaviour of other SW~Sex stars where the Balmer lines 
show only a shallow eclipse compared to the continuum.
\end{abstract} 


\section{Introduction}
\label{sec:introduction}

Cataclysmic variables (CVs) are close binary stars consisting of a red dwarf 
secondary transferring material onto a white dwarf primary via an accretion 
disc or magnetic accretion stream. AC~Cnc and V363~Aur are both examples of 
nova--likes (NLs), defined as CVs which have 
never been observed to undergo nova or dwarf--nova type outbursts.
See \scite{warner95a} for a comprehensive review of CVs.

A knowledge of the masses of the component stars in CVs is 
fundamental to our understanding of the origin, evolution and behaviour of 
these systems. Population synthesis models and the disrupted magnetic 
braking model of CV evolution can be observationally tested only if the number 
of reliably known CV masses increases. 
One of the most reliable ways to measure the masses of CVs is to use the radial 
velocity and the rotational broadening of the secondary star in eclipsing 
systems; the radial velocity of the disc emission lines may not represent 
the white dwarf's orbital motion. At present, reliable masses are 
known for only $\sim$20 CVs, partly due to the difficulties in measurement (see 
\pcite{smith98} for a review).

AC~Cnc was classified as an eclipsing variable star by \scite{kurochkin80} 
with an orbital period of 7.2 hours. \scite{shugarov81} suggested that 
AC~Cnc is a NL based on $UBV$ colours, and the CV nature of AC~Cnc was 
confirmed spectroscopically by \scite{okazaki82} through broad H and He 
emission lines and the later study of \scite{yamasaki83}. 
\scite{downes82} discovered secondary star features in the spectra leading 
to the first mass determination by \scite{schlegel84}, who found 
$M_1$ = 0.82 $\pm$ 0.13 $M_\odot$ and $M_2$ = 1.02 $\pm$ 0.14 $M_\odot$ 
from the radial velocities of the primary and secondary components.

V363~Aur (= Lanning 10) was discovered by \scite{lanning73} as a UV--bright 
source and later found to be a CV through broad Balmer and HeII 
emission by \scite{margon81}, and its typical CV energy distribution 
(\pcite{szkody81}). \scite{horne82} obtained the first spectroscopic and 
photometric data, finding that V363~Aur is an eclipsing system with a 
period of 7.7 hours. \scite{schlegel86} calculated the component masses of 
V363~Aur to be $M_1$ = 0.86 $\pm$ 0.08 $M_\odot$ and 
$M_2$ = 0.77 $\pm$ 0.04 $M_\odot$ from the radial velocities of the 
HeII $\lambda$4686\AA$\:$ emission line and the $G$--band absorption.

The existing component masses of AC~Cnc and V363~Aur use emission line 
radial velocity curves that exhibit phase shifts, suggesting that they 
could be unreliable. In addition to this, the mass ratio of AC~Cnc 
found by \scite{schlegel84} of $q$ = 1.24 $\pm$ 0.08 is the highest known 
of any CV and is very close to the upper limit of mass transfer stability 
computed by the models of \scite{politano96}. 
In this study, we present photometry and spectroscopy 
of AC~Cnc and V363~Aur to calculate new masses using the secondary 
star properties alone.

\section{Observations and Reduction}
\label{sec:observations}

On the nights of 2001 January 9--14 we obtained blue and red 
spectra of AC Cnc and V363 Aur with the 2.5-m Isaac Newton 
Telescope (INT) + IDS spectrometer on La Palma. The blue setup comprised of the 
235-mm camera with the R1200B grating and the EEV10 CCD chip, which gave a 
wavelength coverage of approximately 4490--5580\AA$\:$ at 
0.95-\AA$\:$ (57 km\,s$^{-1}$) resolution. In the red we used the 500-mm 
camera with the R1200Y grating and the TEK5 CCD chip resulting in a 
wavelength range of 6320--6720\AA$\:$ at a resolution of 
0.8-\AA$\:$ (36 km\,s$^{-1}$). Simultaneous photometry in the 
Johnson--Cousins $B$ and $R$ bands was recorded with the 1-m 
Jacobus Kapteyn Telescope (JKT) using 
the SITe2 CCD chip. Full phase coverage was achieved 
for both objects -- a full journal of observations is given in 
Table~\ref{tab:journal}, including the exposure times used.

We also collected 19 spectral type templates ranging from 
G5V--M2V, telluric stars to remove atmospheric features and flux 
standards on both the INT and JKT. Seeing varied between 1.0 and 1.5 
arcsec over the observing run and conditions were photometric on all nights 
except for January 10 when some patchy cloud was present. 

The spectra and images were reduced using standard procedures (e.g. 
\pcite{dhillon94}; \pcite{thoroughgood01}). Comparison arc 
spectra were taken every 40--50 min to calibrate instrumental 
flexure. The arcs were fitted with a sixth--order polynomial in blue and 
a fourth--order polynomial in red with rms scatters of better than 0.01\AA. 
The photometry data were 
corrected for the effects of atmospheric extinction by subtracting the 
magnitude of a nearby comparison star (AC~Cnc--9 and V363~Aur--3; 
\pcite{henden95}) and using values obtained by the CAMC telescope 
(\pcite{helmer85}). 
The absolute photometry is accurate to approximately $\pm$0.5 mJy; the 
relative photometry $\pm$0.03 mag. 
Slit losses were then corrected for by dividing each AC~Cnc and 
V363~Aur spectrum by the ratio of the flux in the spectrum (summed 
over the whole spectral range) to the corresponding photometric flux.


\begin{table*}
{\protect\small
\caption{Journal of observations for AC~Cnc and V363~Aur. The epochs 
are calculated using the new ephemerides presented in this paper 
(equations~\ref{eqn:accnc_ephem} and~\ref{eqn:v363aur_ephem}).}
\label{tab:journal}
\begin{tabular*}{0.96\textwidth}{lcccccccr@{:}lr@{:}lr@{.}lr@{.}l}

\hline
\vspace{-3mm}\\
\multicolumn{1}{c}{UT Date} & \multicolumn{1}{c}{Object}
& \multicolumn{1}{c}{INT} & \multicolumn{1}{c}{No. of} 
& \multicolumn{1}{c}{Exposure}
& \multicolumn{1}{c}{JKT} & \multicolumn{1}{c}{No. of}
& \multicolumn{1}{c}{Exposure}
& \multicolumn{2}{c}{UT} & \multicolumn{2}{c}{UT}
& \multicolumn{2}{c}{Epoch} & \multicolumn{2}{c}{Epoch}\\

\multicolumn{1}{c}{} & \multicolumn{1}{c}{}
& \multicolumn{1}{c}{setup} & \multicolumn{1}{c}{spectra} 
& \multicolumn{1}{c}{time (s)}
& \multicolumn{1}{c}{filter} & \multicolumn{1}{c}{images}
& \multicolumn{1}{c}{time (s)} 
& \multicolumn{2}{c}{start} & \multicolumn{2}{c}{end}
& \multicolumn{2}{c}{start} & \multicolumn{2}{c}{end}\\
\vspace{-3mm}\\
\hline
\vspace{-3mm}\\
2001 Jan 09 & V363 Aur & Red & 61 & 300 & $R$ & 323 & 30 & 22&35 & 03&55 
& 22915&74 & 22916&43\\
2001 Jan 10 & V363 Aur & Blue & 39 & 400 & $B$ & 272 & 30 & 20&07 & 04&36 
& 22918&53 & 22919&63\\
2001 Jan 11 & V363 Aur & Blue & 70 & 400 & $B$ & 474 & 30 & 19&33 & 04&06 
& 22921&57 & 22922&67\\
2001 Jan 12 & V363 Aur & Red & 64 & 300 & $R$ & 348 & 30 & 22&32 & 04&11 
& 22925&97 & 22926&79\\
2001 Jan 12 & AC Cnc & Red & 33 & 300 & $R$ & 179 & 30 & 04&26 & 07&21 & 
--6&14 & --5&74\\
2001 Jan 13 & AC Cnc & Red & 96 & 300 & $R$ & 530 & 30 & 22&39 & 07&04 & 
--3&31 & --2&45\\
2001 Jan 14 & AC Cnc & Blue & 90 & 300 & $B$ & 521 & 30 & 22&27 & 06&33 & 
--0&32 & 0&80\\
\vspace{-3mm}\\
\hline
\end{tabular*}
}
\end{table*}

\section{Results}
\subsection{Ephemeris}
\label{sec:ephem}

We derived new ephemerides for AC~Cnc and V363~Aur, which are used to phase 
all data presented in this paper.
The times of mid--eclipse were determined by 
fitting a parabola to the eclipse minima in the JKT data. 

In the case of AC~Cnc, a least--squares fit to the 29 eclipse timings 
listed in Table~\ref{tab:o-cs} (a) yields the ephemeris:
\begin{equation}
\label{eqn:accnc_ephem}
\begin{array}{lr@{.}llr@{.}ll}
T_{\rm mid-eclipse} = & \!\!\!\! {\rm HJD}\,\,2\,451\,924&5368 
& \!\!\!\! + \!\!\!\! & 0&30047747 & \!\!\!\!\! E \\
& \!\! \pm \,\, 0&0006 & \!\!\!\! \pm \!\!\!\! & 0&00000004. & \\
\end{array}
\end{equation}
A least--squares fit to the 17 eclipse timings of V363~Aur listed in 
Table~\ref{tab:o-cs} (b) gives the ephemeris:
\begin{equation}
\label{eqn:v363aur_ephem}
\begin{array}{lr@{.}llr@{.}ll}
T_{\rm mid-eclipse} = & \!\!\!\! {\rm HJD}\,\,2\,444\,557&9514 
& \!\!\!\! + \!\!\!\! & 0&32124187 & \!\!\!\!\! E \\
& \!\! \pm \,\, 0&0016 & \!\!\!\! \pm \!\!\!\! & 0&00000008. & \\
\end{array}
\end{equation}
We see no evidence for any systematic variation in the O--C values shown in 
Table~\ref{tab:o-cs} in either AC~Cnc or V363~Aur.

\begin{table}[h]
{\protect\small
\caption{(a) Times of mid--eclipse for AC~Cnc according to 
Kurochkin \& Shugarov (1980; KS80), Yamasaki, Okazaski \& Kitamura 
(1983; YOK83), Schlegel, Kaitchuck \& Honeycutt (1984; SKH84), 
Zhang (1987; Z87) and this paper. 
(b) Times of mid--eclipse for V363~Aur according to Horne, 
Lanning \& 
Gomer (1982; HLG82), Schlegel, Honeycutt \& Kaitchuck (1986; SHK86), Rutten, 
van Paradijs \& Tinbergen (1992; RvPT92) and this paper. 
The uncertainties on the mid--eclipse times measured from our data are 
0.001; all other uncertainties are taken to be 0.005.}
\label{tab:o-cs}
\begin{tabular*}{77mm}{lr@{.}lr@{.}lc}

(a) AC~Cnc\\
\\
\hline
\vspace{-3mm}\\
\multicolumn{1}{c}{Cycle} & 
\multicolumn{2}{c}{HJD}
& \multicolumn{2}{c}{O--C} & 
\multicolumn{1}{c}{Reference} \\
\multicolumn{1}{c}{(E)} & 
\multicolumn{2}{c}{at mid--eclipse}
& \multicolumn{2}{c}{(secs)} & \multicolumn{1}{c}{}\\
\multicolumn{1}{c}{} & 
\multicolumn{2}{c}{(2,400,000+)}
& \multicolumn{2}{c}{} & 
\multicolumn{1}{c}{} \\
\vspace{-3mm}\\
\hline
\vspace{-3mm}\\
--59456 & 34059&348 & --37&03 & KS80\\
--50866 & 36640&446 & --336&27 & KS80\\
--50813 & 36656&367 & --708&29 & KS80\\
--47342 & 37699&327 & --474&74 & KS80\\
--33860 & 41750&380 & 886&55 & KS80\\
--31241 & 42537&320 & --20&02 & KS80\\
--30136 & 42869&344 & --331&40 & KS80\\
--29177 & 43157&504 & --149&38 & KS80\\
--29074 & 43188&450 & --424&07 & KS80\\
--29061 & 43192&362 & 76&43 & KS80\\
--29038 & 43199&275 & 250&81 & KS80\\
--28848 & 43256&368 & 447&87 & KS80\\
--28785 & 43275&293 & 8&90 & KS80\\
--28165 & 43461&579 & --857&79 & KS80\\
--26867 & 43851&619 & 891&34 & KS80\\
--26598 & 43932&439 & 162&19 & KS80\\
--26588 & 43935&443 & 95&25 & KS80\\
--26578 & 43938&448 & 114&72 & KS80\\
--26525 & 43954&376 & 347&49 & KS80\\
--23098 & 44984&1119 & 313&96 & YOK83\\
--23094 & 44985&3141 & 339&02 & YOK83\\
--23088 & 44987&1165 & 298&86 & YOK83\\
--21522 & 45457&6637 & 254&17 & SKH84\\
--19342 & 46112&7032 & 134&62 & Z87\\
--19335 & 46114&8053 & 27&29 & Z87\\
--19312 & 46121&7177 & 149&82 & Z87\\
--6 & 51922&7339 & --8&03 & This Paper\\
--3 & 51923&6350 & --36&75 & This Paper\\
$\:$0 & 51924&5367 & --13&63 & This Paper\\
\hline\\
(b) V363~Aur
\\
\\
\hline
\vspace{-3mm}
\\
0 & 44557&9495 & --168&03 & HLG82\\
3 & 44558&9128 & --204&81 & HLG82\\
6 & 44559&8772 & --146&54 & HLG82\\
105 & 44591&6813 & --46&80 & HLG82\\
106 & 44592&0023 & --67&70 & HLG82\\
4694 & 46065&8614 & 51&42 & SHK86\\
4700 & 46067&7877 & --48&05 & SHK86\\
4703 & 46068&7544 & 208&94 & SHK86\\
4706 & 46069&7182 & 215&37 & SHK86\\
9365 & 47566&3817 & 8&49 & RvPT92\\
9368 & 47567&3466 & 109&96 & RvPT92\\
10057 & 47788&6818 & 70&93 & RvPT92\\
10060 & 47789&6457 & 85&99 & RvPT92\\
10082 & 47796&7142 & 187&84 & RvPT92\\
10088 & 47798&6401 & 53&81 & RvPT92\\
22916 & 51919&5301 & --12&64 & This Paper\\
22922 & 51921&4577 & 0&21 & This Paper \\
\vspace{-3mm}\\
\hline

\end{tabular*}
}
\end{table}

\subsection{Average Spectrum}
\label{sec:average}

\begin{figure*}
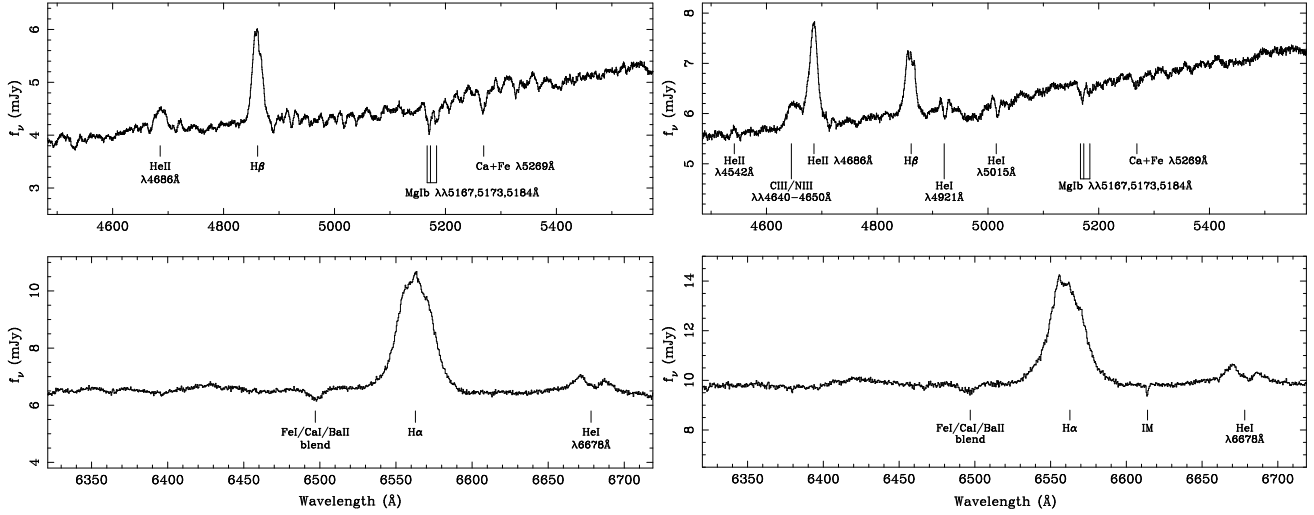

\begin{center}
\begin{tabular}{cc}
\includegraphics[width=6.75cm,angle=-90]{accnc_average.ps}
\includegraphics[width=6.75cm,angle=-90]{v363aur_average.ps}
\end{tabular}
\end{center}
\caption{\protect\small
Average spectra of AC Cnc (left) and V363 Aur (right).}
\label{fig:average}
\end{figure*}

The average spectra of AC~Cnc and V363~Aur are shown in 
Fig.~\ref{fig:average}, and in Table~\ref{tab:linewidths} we list fluxes, 
equivalent widths (EW) and velocity widths of the most prominent lines measured 
from the average spectra.

Both systems show broad, symmetric, single--peaked Balmer emission lines 
instead 
of the double--peaked profiles one would expect from a high inclination 
accretion disc, much like other nova--like systems (e.g. \pcite{warner95a}). 
The HeI lines, however, are broad and double--peaked in nature. 
The line strength of 
HeII $\lambda$4686\AA$\:$ is much stronger in V363~Aur than AC~Cnc and even 
more dominant than H$\beta$ emission. Another high excitation feature, 
the CIII/NIII $\lambda\lambda$4640--4650\AA$\:$ blend, is only present 
in V363~Aur and is very broad.
The emission lines are characteristic of the SW~Sex stars (e.g. 
\pcite{dhillon97b}), but unlike others in this sub--class, these systems 
show clear secondary star features (no doubt due to their longer periods and 
hence earlier type secondaries). Both AC~Cnc and V363~Aur show the 
absorption features of the neutral metals CaI, FeI and MgI, even in the 
average spectrum shown in Fig.~\ref{fig:average}, which has not been 
corrected for orbital motion. The secondary star features appear to be 
stronger relative to the continuum in AC~Cnc than in V363~Aur.
The weak feature at $\lambda$6614\AA$\:$ in the V363~Aur red spectrum is an 
interstellar absorption line.

\begin{table*}[h]
{\protect\small
\caption[Fluxes and widths of prominent lines in V363 Aur]{Fluxes and 
widths of prominent lines in AC~Cnc and V363~Aur, measured from the average 
spectra. In the case of V363~Aur, HeII$\:\lambda$4686\AA$\:$ and 
CIII/NIII$\:\lambda\lambda$4640--4650\AA$\:$ are blended, so separate values 
of the flux and EW are given (determined from a double--Gaussian fit) as 
well as the combined flux of the two.}
\label{tab:linewidths}
\begin{tabular*}
{0.665\textwidth}{
lr@{$\:\pm\:$}lr@{$\:\pm\:$}lr@{$\:\pm\:$}lr@{$\:\pm\:$}l}

(a) AC~Cnc\\
\\
\hline
\vspace{-3mm}\\
\multicolumn{1}{l}{Line} & 
\multicolumn{2}{c}{Flux} & 
\multicolumn{2}{c}{EW} & 
\multicolumn{2}{c}{FWHM} &
\multicolumn{2}{c}{FWZI}\\
\multicolumn{1}{c}{} & 
\multicolumn{2}{c}{$\times$ 10$^{-14}$}
& \multicolumn{2}{c}{(\AA)} & \multicolumn{2}{c}{(km\,s$^{-1}$)} 
& \multicolumn{2}{c}{(km\,s$^{-1}$)}\\
\multicolumn{1}{c}{} & 
\multicolumn{2}{c}{(ergs\,cm$^{-2}$\,s$^{-1}$)}
& \multicolumn{2}{c}{} & \multicolumn{2}{c}{} 
& \multicolumn{2}{c}{}\\
\vspace{-3mm}\\
\hline
\vspace{-3mm}\\
H$\alpha$ & 7.57&0.01 & 16.55&0.03 & 1100&100 & 
2900&500  \\

H$\beta$ & 4.63&0.01 & 8.57&0.03 & 1250&100 & 
2800&300  \\

HeI$\:\lambda$6678\AA & 0.94&0.01 & 2.17&0.02 & 1100&100 & 
1900&500  \\

HeII$\:\lambda$4686\AA & 1.36&0.01 & 2.32&0.02 & 1600&100 & 
2300&300  \\
\hline\\
(b) V363~Aur
\\
\\
\hline
\vspace{-3mm}
\\
H$\alpha$ & 8.07&0.02 & 11.82&0.02 & 1150&100 & 
3100&500  \\

H$\beta$ & 3.60&0.02 & 4.88&0.02 & 1250&100 & 
2900&300  \\

HeI$\:\lambda$6678\AA & 0.86&0.01 & 1.31&0.02 & 1100&100 & 
2300&500  \\

HeII$\:\lambda$4686\AA & 4.84&0.02 & 6.19&0.03 & 1150&100 & 
3400&500  \\

CIII/NIII$\:\lambda\lambda$4640--4650\AA & 2.05&0.03 & 2.58&0.03 
& 1950&100 & 4900&500  \\

HeII + CIII/NIII & 7.35&0.02 & 9.37&0.03 & 
\multicolumn{2}{c}{} & \multicolumn{2}{c}{} \\
\vspace{-3mm}\\
\hline
\end{tabular*}
}
\end{table*}

\subsection{Light Curves}
\label{sec:light}

\begin{figure*}
\begin{center}
\begin{tabular}{c}
\multicolumn{1}{l}{(a) AC~Cnc}\\
\includegraphics[width=10.5cm,angle=-90]{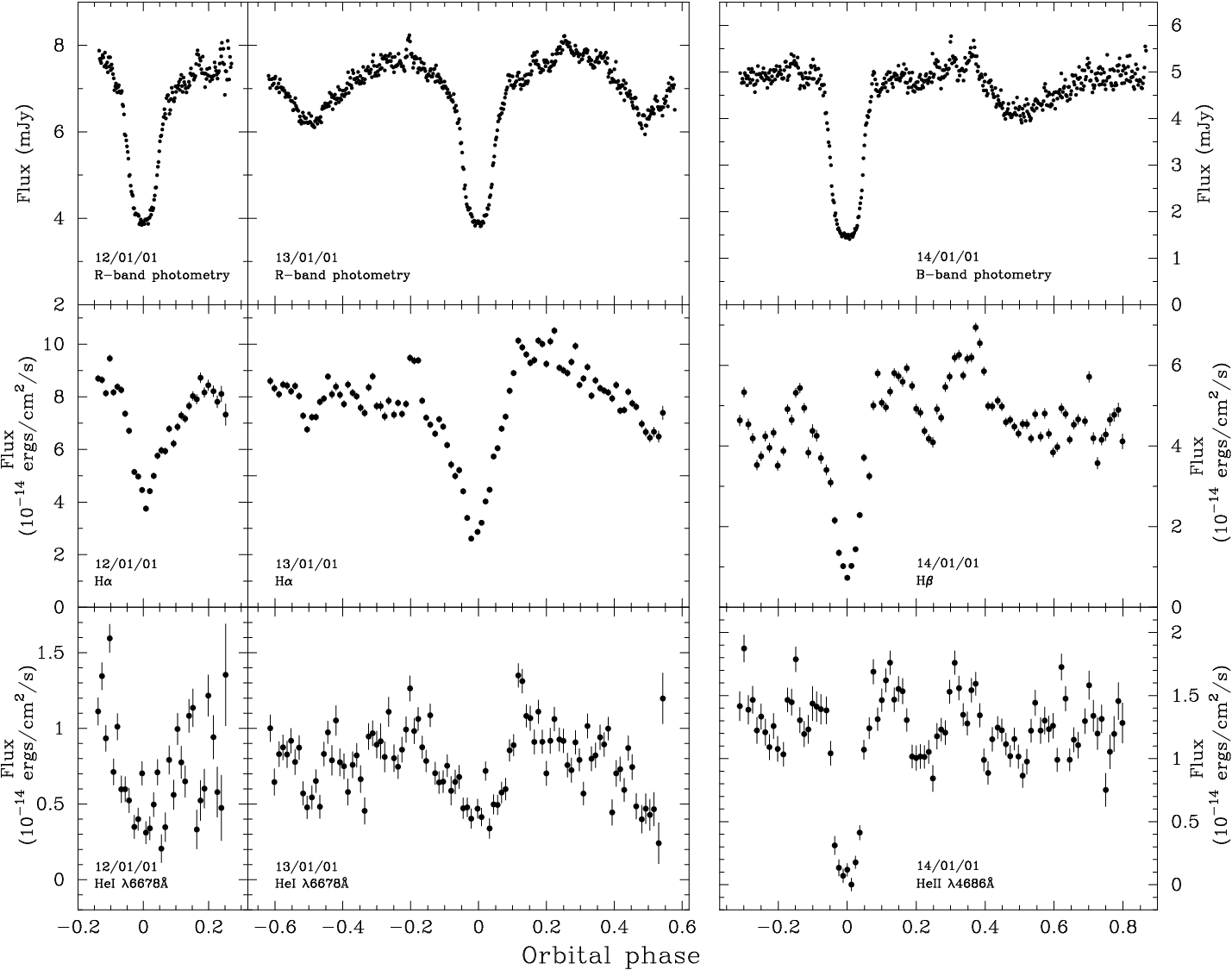}\\
\multicolumn{1}{l}{(b) V363~Aur}\\
\includegraphics[width=10.1cm,angle=-90]{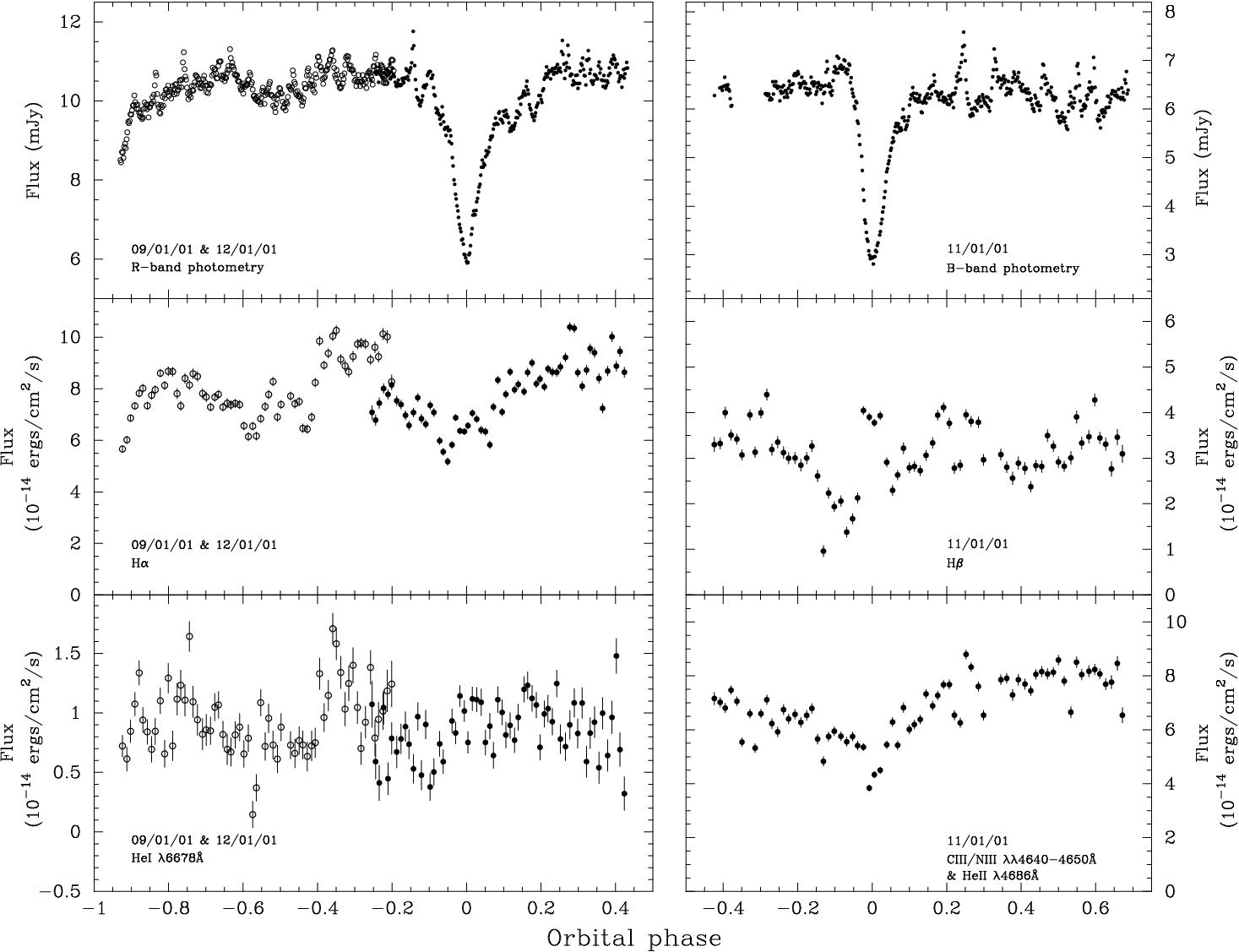}\\
\end{tabular}
\end{center}
\caption{\protect\small
Lightcurves of (a) AC~Cnc and (b) V363~Aur. Top panels: $R$ and 
$B$--band JKT lightcurves. 
Middle and bottom panels: Emission--line lightcurves of H$\alpha$, 
H$\beta$, CIII/NIII~$\lambda\lambda$4640--4650\AA$\:$ + 
HeII~$\lambda$4686\AA$\:$ and 
HeI~$\lambda$6678\AA. The $R$--band V363~Aur data are from 2 nights; open 
circles represent data from 12/01/01 and closed circles represent data 
from 09/01/01. Note that the two nights of data are split by orbital phase, 
as well as by night of observation.}
\label{fig:lc}
\end{figure*}

The top panels of Fig.~\ref{fig:lc}(a) and ~\ref{fig:lc}(b) show the $B$ 
and $R$--band JKT light curves. The remaining panels show 
emission--line light curves, which were produced by subtracting a polynomial 
fit to the continuum and summing the residual flux. All
light curves are plotted as a function of phase following our new ephemerides. 

The $B$ and $R$--band JKT light curves of AC~Cnc show deep, symmetrical primary 
eclipses with out-of-eclipse magnitudes of 14.30 $\pm$ 0.05 mag in $B$ and 
14.00 $\pm$ 0.05 in $R$. The primary eclipse depth is 1.8 mag in $B$ and 
0.73 mag in $R$. We measured the phase half--width of eclipse at the 
out-of-eclipse level ($\Delta\phi$) by timing the first and last contacts 
of the $B$ and $R$--band photometry eclipses and dividing by two. 
Our average value of $\Delta\phi$ = 0.09 $\pm$ 0.01 phases is smaller than, but 
consistent with, the value of 0.108 $\pm$ 0.008 quoted in \scite{harrop96}.
Ellipsoidal modulation of the red star is clearly present, although flaring 
around phases 0.3--0.4 contaminates the effect in $B$. There is 
also evidence for a secondary eclipse at phase 0.5. We see no evidence for 
a bright--spot in the light curves but flickering is present, particularly 
just after primary eclipse. A notable feature of the $B$ and $R$--band JKT 
light curves is the U--shaped eclipse minima, in contrast to the V--shaped 
minima seen in many SW~Sex systems (\pcite{knigge00}). The eclipses of the 
Balmer lines show the usual V--shape, but the HeII line has a U--shaped 
eclipse minimum and is completely eclipsed, suggesting an origin close to 
the white dwarf. The H$\alpha$ flux 
increases markedly after eclipse before slowly declining -- there is also 
the suggestion of a sharp decrease in flux around phase 0.5. This 
secondary eclipse is possibly also seen in the HeI line, 
although the primary eclipse here is much broader and shallower. The H$\beta$ 
flux seems more erratic in behaviour, closely resembling the 
higher--excitation HeII emission line. Note that when the Balmer flux increases, 
the flickering in the JKT light curves is more prominent.

The V363~Aur JKT light curves in both the $B$ and $R$--bands are deep and 
symmetrical with V--shaped eclipse minima, much like the SW~Sex systems 
(\pcite{knigge00}). The $B$--band out-of-eclipse magnitude is 14.50 $\pm$ 
0.10, and the eclipse depth is 0.88 mag; in the $R$--band, 
the out-of-eclipse magnitude is 13.65 $\pm$ 0.05 with an eclipse depth of 
0.63 mag. Our measured phase half--width of eclipse at the out-of-eclipse 
level, $\Delta\phi$ = 0.078 $\pm$ 0.005 (the average from the $B$ and 
$R$--band photometry eclipses) is lower than the value of 0.120 $\pm$ 0.010 
quoted by \scite{harrop96}. 
There is evidence of either a shallow secondary eclipse or orbital modulation 
in the $R$--band light curve but not so in $B$. One of the most notable features 
of the light curves is the high level of flickering present. 
The red emission--line light curves of H$\alpha$ and HeI are similar in that 
they both show a maximum flux around phase 0.7. Perhaps the most 
interesting feature of the emission--line lightcurves is in the primary 
eclipse of the Balmer lines; the flux seems to drop entering eclipse but at 
phase 0 there in a sharp increase in Balmer line emission followed by a 
rapid decrease. This is particularly prominent in H$\beta$, but also seems to be 
present in the H$\alpha$ line and possibly the HeI line. The effect is 
definitely not present in the high excitation HeII and CIII/NIII complex.

\subsection{Trailed spectrum \& Doppler Tomography}
\label{sec:trailed_spectrum}

\begin{figure*}
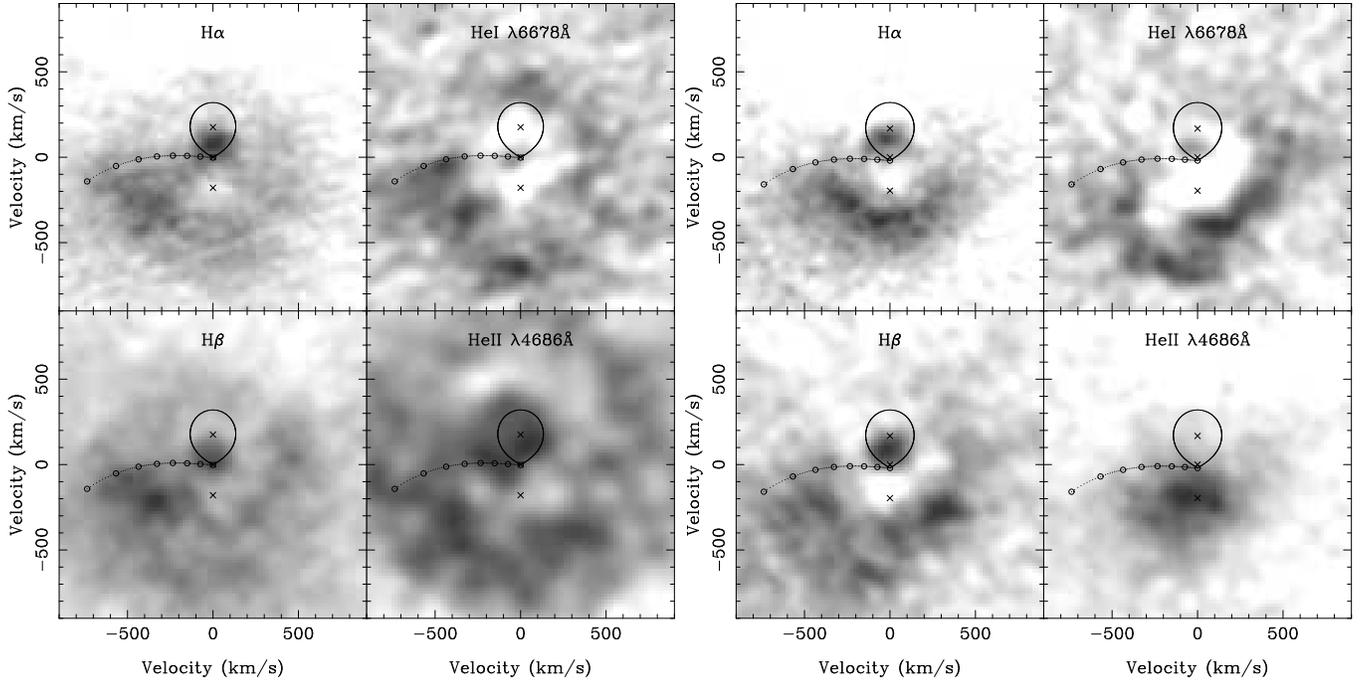

\begin{center}
\begin{tabular}{cc}
\includegraphics[width=9.0cm,angle=-90]{accnc_dopplermap_irr.ps}
\includegraphics[width=9.0cm,angle=-90]{v363aur_dopplermap_irr.ps}
\end{tabular}
\end{center}
\caption{\protect\small
Doppler maps of emission lines in AC~Cnc (left) and V363 Aur 
(right). The three crosses marked on each Doppler map 
represent the centres of mass of the secondary (upper cross), the system 
(middle cross) and the primary (lower cross). These crosses, the Roche lobe 
of the secondary star and the predicted trajectory of the gas stream have 
been plotted using the $K_R$--corrected system parameters summarised in 
Table~\ref{tab:params}. The series of circles 
along the gas stream mark the distance from the white dwarf at intervals 
of 0.1$L_1$, ranging from 1.0$L_1$ at the red star to 0.3$L_1$.}
\label{fig:doppler}
\end{figure*}

We subtracted polynomial fits to the continuum from the spectra and 
then rebinned the spectra onto a constant velocity--interval scale 
centred on the rest wavelength of the H$\alpha$, H$\beta$, 
HeI $\lambda$6678\AA$\:$ and HeII $\lambda$4686\AA$\:$ lines. The data were 
then phase--binned into 50 bins, all of which were filled except for 1 empty 
bin at phase 0.31--0.33 in the blue V363~Aur data set. 
The trailed spectra of the 
lines are shown in the upper panels of Fig.~\ref{fig:trailed} (a) and (b). 
We then constructed Doppler tomograms from the trailed spectra, a technique 
which maps the velocity--space distribution of the emission lines 
(e.g. \pcite{marsh00}). The Doppler maps are shown in Fig.~\ref{fig:doppler} 
and trailed spectra reconstructed from the maps are presented in the lower 
panels of Fig.~\ref{fig:trailed} (a) and (b). 

The trailed spectra of AC~Cnc show two clearly--defined components. 
The first is a high--amplitude S--wave with a semi--amplitude 
of $\sim$500km\,s$^{-1}$, which crosses zero velocity from red to 
blue around phase 0.15. The emission is particularly noticable in H$\alpha$ 
but can be seen in the other lines. This component appears in the Doppler map 
superimposed upon a ring of emission characteristic of an accretion disc 
at $R_D = 0.4-0.5 L_1$. This is what one would expect from the bright--spot, 
although it appears slightly downstream from where the computed gas stream 
joins the accretion disc, a trait also seen in other CVs 
(e.g. WZ~Sge; \pcite{spruit98}). 
The second, lower--velocity component seen in the Balmer lines, 
seemingly in anti-phase with the higher--velocity component, shows up clearly 
in the Doppler maps as originating on the inner hemisphere of the secondary 
star. There is possibly another component visible in the 
HeI $\lambda$6678\AA$\:$ line, which could be interpreted as the signature 
of a faint double--peaked accretion disc. All of the low--excitation lines 
exhibit the rotational disturbance 
expected from a high inclination accretion disc, but interestingly this is 
not seen in the HeII $\lambda$4686\AA$\:$ line. These emission features are 
very similar to those seen in the novalike SW~Sex (\pcite{dhillon97b}).

The high--amplitude S--wave and the low--amplitude emission are 
also seen in V363~Aur, although the anti--phase sinusoid of the secondary 
is less well pronounced. The high--amplitude Balmer emission in the 
Doppler maps appears as a bright--spot downstream from where the gas stream 
meets the disc, as well as from the opposite side of the disc. There is 
clearly Balmer emission from the inner hemisphere of the secondary which is 
likely to be caused by irradiation by the accretion regions or white dwarf. 
The HeI trailed spectrum gives the visual impression of an absorption feature 
(moving at $\sim K_W$) super--imposed on a high--velocity S--wave.
The strange increase in Balmer line and 
HeI $\lambda$6678\AA$\:$ emission at phase zero seen in the lightcurves is 
clearly visible in the trailed spectra. Possible explanations for this 
are given in Section~\ref{sec:discussion}. The high-excitation HeII 
$\lambda$4686\AA$\:$ line is single-peaked throughout the orbit and the 
emission lies on the white dwarf centre of mass in the Doppler map. 

\begin{figure*}
\begin{center}
\begin{tabular}{c}
\multicolumn{1}{l}{(a) AC~Cnc}\\
\includegraphics[width=9.9cm,angle=-90]{accnc_trailed+computed.ps}\\
\\
\multicolumn{1}{l}{(b) V363~Aur}\\
\includegraphics[width=10.4cm,angle=-90]{v363aur_trailed+computed.ps}\\
\end{tabular}
\end{center}
\caption{
(a) Trailed spectra of AC~Cnc (upper panel) \& data computed from the 
Doppler maps (lower panel). (b) Trailed spectra of V363~Aur (upper panel) 
\& data computed from the Doppler maps (lower panel). A comparison of the 
upper and lower panels gives some indication of the quality of the fit. 
The gaps in the computed data correspond to eclipse spectra (selected 
using the light curves of Fig.~\ref{fig:lc}), which have been 
omitted from the fit as Doppler tomography cannot properly account 
for these phases. Note that the data have been folded to show more than one 
binary cycle.
}
\label{fig:trailed}
\end{figure*}

\subsection{Radial velocity of the white dwarf}
\label{sec:whitedwarf}

We measured the radial velocities of the emission lines in AC~Cnc and 
V363~Aur by applying the double--Gaussian method of 
\scite{schneider80}, since this technique is sensitive mainly to the 
line wings and should therefore reflect the motion of the white dwarf with 
the highest reliability. We used Gaussians of widths 200, 300 and 
400 km\,s$^{-1}$ and varied their separation from 200 to 2500 
km\,s$^{-1}$. We then fitted 
\begin{equation}
V=\gamma-K\sin[2\pi(\phi-\phi_0)]
\end{equation}
to each set of measurements, where $V$ is the radial velocity, $K$ the 
semi--amplitude, $\phi$ the orbital phase, and $\phi_0$ is the phase at which 
the radial velocity curve crosses from red to blue. 

\begin{figure*}
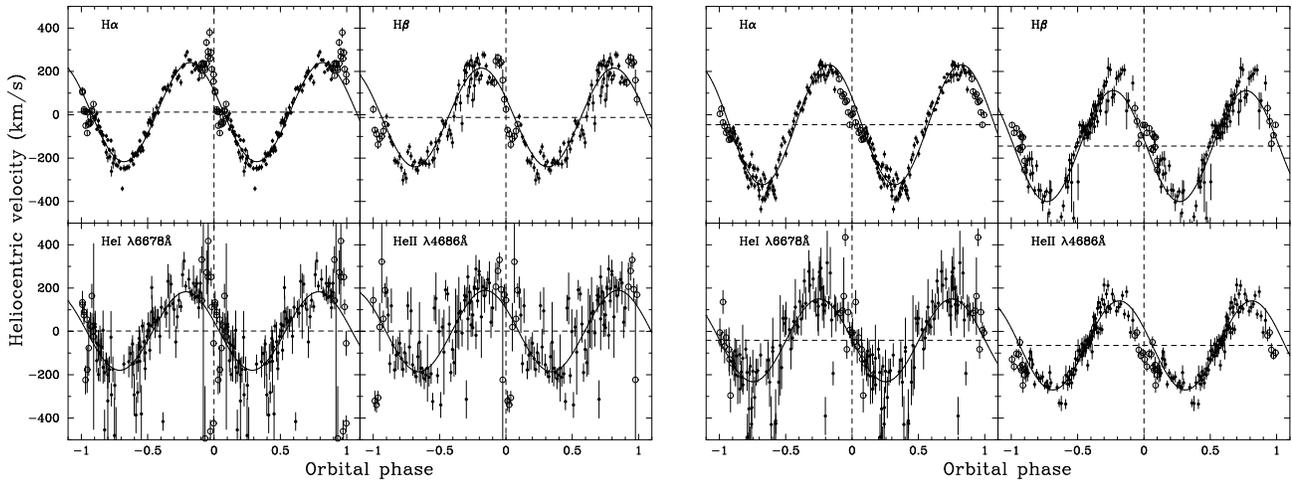

\begin{center}
\begin{tabular}{ccc}
\includegraphics[width=6.3cm,angle=-90]{accnc_radial2.ps}
\hspace{5mm}
\includegraphics[width=6.3cm,angle=-90]{v363aur_radial2.ps}
\end{tabular}
\end{center}
\caption{\protect\small
Radial velocity curves of the emission lines in AC Cnc using Gaussian 
widths of 200 km\,s$^{-1}$ and a separation of 1200 km\,s$^{-1}$ (left), 
and of V363 Aur using Gaussian widths of 300 km\,s$^{-1}$ and a separation 
of 1200 km\,s$^{-1}$(right). 
We omitted the points around primary eclipse (open circles) during the 
fitting procedure as these measurements are affected by the 
rotational disturbance.
The horizontal dashed lines represent the $\gamma$ velocities given by 
the sinusoidal fits.
}
\label{fig:rvs}
\end{figure*}

Examples of the radial velocity curves for AC~Cnc and V363~Aur are shown in 
Fig.~\ref{fig:rvs}. The most striking feature of all of the radial velocity 
curves are the phase shifts, where the spectroscopic conjunction of each line 
occurs after photometric eclipse. This phase shift implies an emission line 
source trailing the accretion disc, such as a bright spot, 
and is a common feature of SW~Sex stars (e.g. DW~UMa, \pcite{shafter88}; 
V1315~Aql, \pcite{dhillon91}; SW~Sex, \pcite{dhillon97b}). 
There is clear evidence of rotational disturbance in the Balmer lines of 
AC~Cnc, where the radial velocities measured just prior to eclipse are 
skewed to the red, and those measured after eclipse are skewed to the blue. 
This confirms the detection of a similar feature in the trailed spectra, and 
indicates that at least some of the emission must originate in the disc. 
We tried to measure white dwarf radial velocity ($K_W$) values from the 
emission lines in AC~Cnc 
and V363~Aur using a diagnostic diagram (e.g. \pcite{shafter86}) 
and a light centres diagram (e.g. \pcite{marsh88a}) but with no success.
This is not surprising given that the Doppler maps (Fig.~\ref{fig:doppler}) 
show that the accretion disc does not dominate the emission in these systems.
We conclude that the emission lines of AC~Cnc and V363~Aur are unreliable 
indicators of the white dwarf radial velocity due to the phase shifts.

\subsection{Radial velocity of the secondary star}
\label{sec:secondary}

The secondary star in both AC~Cnc and V363~Aur is clearly visible in 
Fig.~\ref{fig:average} through absorption lines of MgI, FeI and CaI. 
We compared regions of the spectra rich in absorption lines with 
a number of red dwarfs of spectral types G5V--M2V. A technique 
known as skew mapping was used to enhance the secondary features and obtain a 
radial velocity ($K_R$) measurement (see \scite{vandeputte03} for a detailed 
critique of the method and \scite{sad98b} for a successful application to 
BT~Mon). 

The first step was to shift the spectral type template stars to correct for 
their radial velocities. We then normalized each spectrum by dividing 
by a first--order polynomial fit, and then subtracting a higher order fit to 
the continuum. This ensures that line strength is preserved along the 
spectrum. The AC~Cnc and V363~Aur spectra were normalized in the same way. 
The template spectra were artificially broadened to account for the 
orbital smearing of the CV spectra due to their exposure times ($t_{exp}$) 
using the formula
\begin{equation}
{V} =  {{{t_{exp}}{2\pi}{K_R}} \over {P}}
\label{eqn:smear}
\end{equation}
(e.g. \pcite{watson00}), and then by the rotational velocity of the 
secondary ($v \sin i$). Estimated values of $K_R$ and $v \sin i$ were used 
in the first instance, before iterating to find the best--fitting values given 
in Section~\ref{sec:params}.
Regions of the spectrum devoid of emission lines were then cross--correlated 
with each of the templates yielding a time series of cross--correlation 
functions (CCFs) for each template star. To produce the skew maps, these CCFs 
were back--projected in the same way as time--resolved spectra in standard 
Doppler tomography (\pcite{marsh88b}). If there is a detectable 
secondary star, we expect a peak at (0,$K_R$) in the skew map.
This can be repeated for each of the templates, and the final skew map is the 
one that gives the strongest peak.

\begin{figure*}
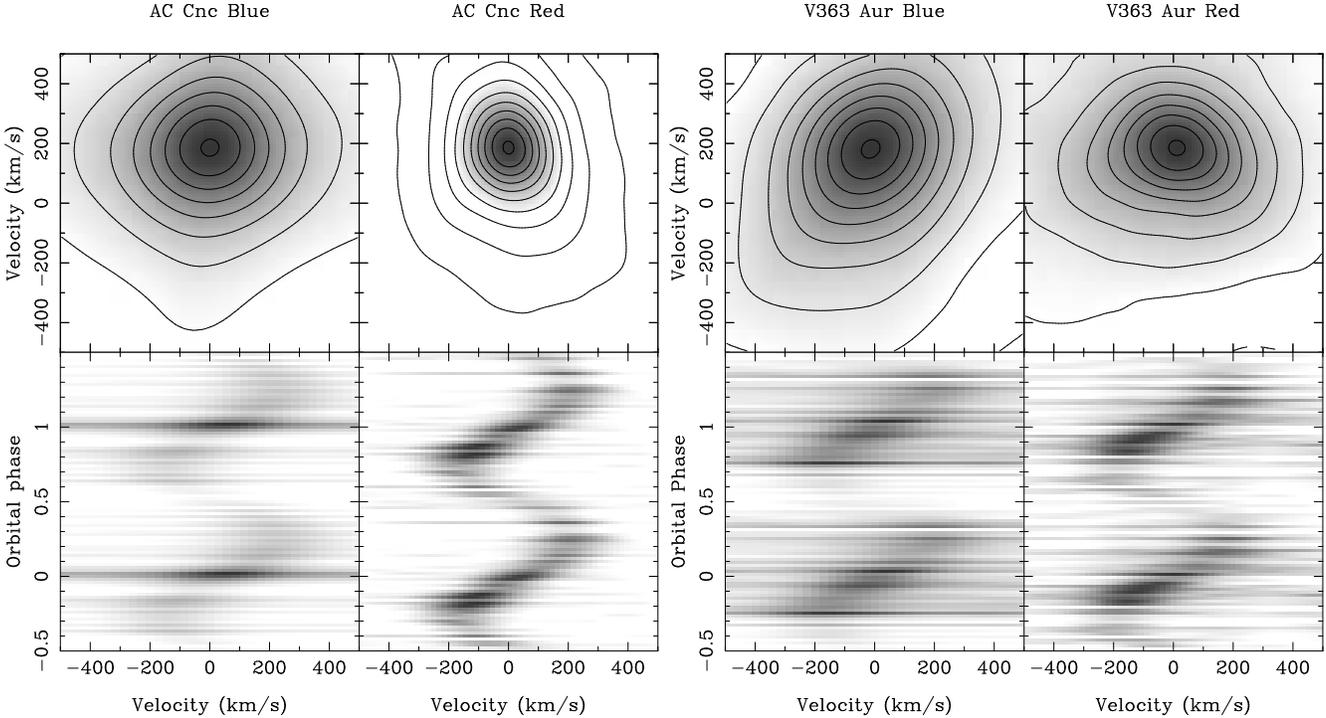

\begin{center}
\begin{tabular}{ll}
\includegraphics[height=9.5cm,angle=0]{accnc_skew+sec.ps}
\includegraphics[height=9.5cm,angle=0]{v363aur_skew+sec.ps}
\end{tabular}
\end{center}
\caption{\protect\small
Left: Skewmaps and trailed CCFs of AC Cnc, measured using 
a K2V template. Right: Skewmaps and trailed CCFs of V363 Aur, 
measured using a G7V template in the blue and a G5V template in the red.}
\label{fig:skewmaps}
\end{figure*}

The AC~Cnc skew maps show well--defined peaks at 
$K_y \approx$ 186 km\,s$^{-1}$ -- the skew map for the K2V template is 
shown in Fig.~\ref{fig:skewmaps} together with the trailed CCFs and 
the regions used for the 
cross--correlation can be seen in Fig.~\ref{fig:residual}. 
A systemic velocity of $\gamma$ = 40 km\,s$^{-1}$ was applied in order to 
shift the skew map peaks onto the $K_x$ = 0 axis. 
The $K_R$ value varies little with $\gamma$ in practice, as 
$K_x \ll K_y$ in the back--projections (e.g. \pcite{sad98b}).
We adopt $\gamma$ = 40 $\pm$ 5 km\,s$^{-1}$ as the 
systemic velocity of the AC~Cnc, in contrast to values of 122 km\,s$^{-1}$ 
and 107 km\,s$^{-1}$ found by \scite{schlegel84} using Balmer emission lines.
Our adopted $K_R$ value of 186 $\pm$ 3 km\,s$^{-1}$ is derived from 
the best--fitting template (K2V), with the error incorporating the 
spread of values obtained by using different templates 
(see Table~\ref{tab:vsini}). 
The uncertainty also reflects the scatter in the radial velocity 
fits shown in Fig.~\ref{fig:irrplot}. 
Note the remarkable agreement in the ($K_y,K_x$) values obtained from 
the red and blue data sets in Table~\ref{tab:vsini}. The small scatter in 
$K_y$ assures us that $K_R$ is robust, and the small scatter in $K_x$ around 
zero indicates that our assumed $\gamma$ is correct.

The final V363~Aur skew maps for the G7V template (blue) and G5V 
template (red) and trailed CCFs are shown in Fig.~\ref{fig:skewmaps}, 
with the regions used for the cross--correlation marked in 
Fig.~\ref{fig:residual}. 
For V363~Aur, the systemic velocity was less simple to determine, as the 
blue skew map suggested $\gamma$ = --10 km\,s$^{-1}$ and the red 
skew map gave $\gamma$ = 10 km\,s$^{-1}$. We can find no explanation 
for this discrepancy, so adopt a systemic velocity of 
$\gamma$ = 0 $\pm$ 10 km\,s$^{-1}$. \scite{schlegel86} find systemic 
velocities ranging 
from 10 $\pm$ 3 km\,s$^{-1}$ for the HeII $\lambda$4686\AA$\:$ line--wings to 
35 $\pm$ 3 km\,s$^{-1}$ for the G-band absorption line. However, the peaks in 
the skew maps appear consistently at $K_y \approx$ 184 km\,s$^{-1}$. 
Our adopted value of 
$K_R$ = 184 $\pm$ 5 km\,s$^{-1}$ once again acknowledges the uncertainty in 
using different templates and the scatter in the radial velocity curves.

\subsection{Rotational velocity and spectral type of the secondary star}
\label{sec:rotational}

In order to maximise the strength of the secondary features, we averaged the 
orbitally--corrected eclipse spectra of AC~Cnc and V363~Aur.
The spectral--type templates were broadened for smearing due to orbital 
motion as before and rotationally broadened by a range of velocities 
(50--240 km\,s$^{-1}$). We then ran an optimal subtraction routine, which 
subtracts a constant times the normalized template spectrum from the 
normalized, orbitally--corrected CV spectrum, adjusting the constant to 
minimize the residual scatter between the spectra. The scatter is measured 
by carrying out the subtraction and then computing the $\chi^2$ between the 
residual spectrum and a smoothed version of itself. By finding the value of 
rotational broadening that minimizes the $\chi^2$, we obtain an 
estimate of both $v \sin i$ and the spectral type of the secondary star. 
Note that the $v \sin i$ values of the template stars are much lower than the 
instrumental resolution, so do not affect our measurements of $v \sin i$ for the 
secondary star.

The value of $v \sin i$ obtained using this method varies depending on the 
spectral type template, the wavelength region for optimal subtraction, 
the amount of smoothing of the residual spectrum in the calculation of 
$\chi^2$ and the value of the limb--darkening coefficient used in the 
broadening procedure. The values of $v \sin i$ found from the G and K templates in 
the red and blue wavelength ranges, calculated using a 
limb--darkening coefficient of 0.5 and smoothed using a Gaussian of 
FWHM = 15km\,s$^{-1}$, are listed in Table~\ref{tab:vsini}, together 
with the minimum $\chi^2$. 
The optimal subtraction technique also tells us the value of the constant 
by which the template spectra were multiplied, which, for normalized spectra, 
is the fractional contribution of the secondary star to the total light 
in eclipse. These results are also summarised in Table~\ref{tab:vsini}. 

\begin{figure*}
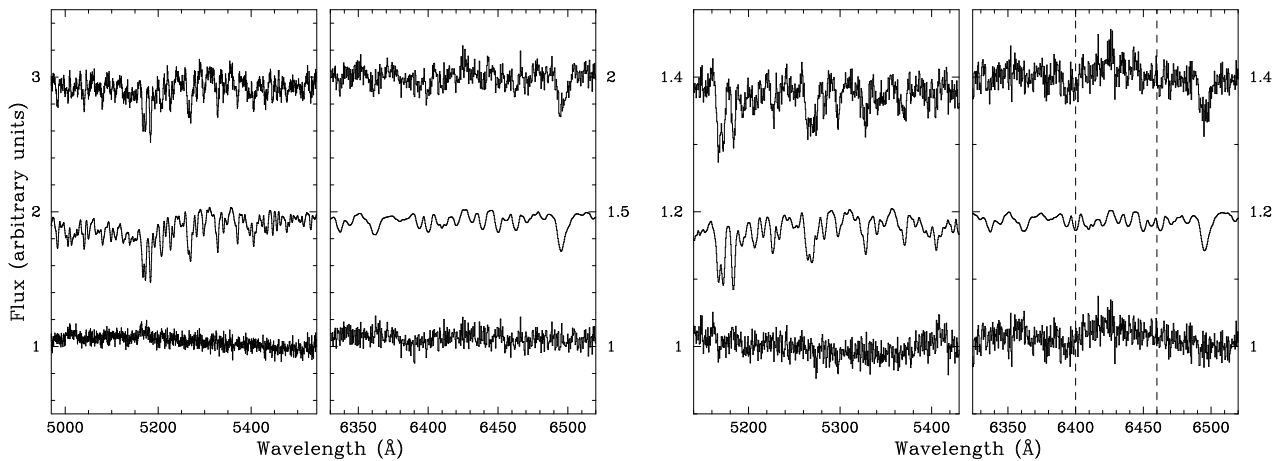

\begin{center}
\begin{tabular}{cc}
\includegraphics[width=6.0cm,angle=-90]{accnc_residual.ps} &
\includegraphics[width=6.0cm,angle=-90]{v363aur_residual.ps}
\end{tabular}

\end{center}
\caption{\protect\small
Orbitally--corrected average eclipse spectra of the CVs (top) with the 
best--fitting broadened template (middle) and the residuals after 
optimal subtraction (bottom). The template spectra have been multiplied by 
the optimal factor found from the optimal subtraction. All of the spectra 
are normalised and have been offset on the plots 
by an arbitrary amount for clarity. Left: AC Cnc (eclipse phases 
$-0.05 < \phi < 0.05$ in blue and red) with a K2V template. 
Right: V363 Aur (eclipse phases $-0.07 < \phi < 0.09$ in blue and 
$-0.09 < \phi < 0.09$ in red) with a G5V (red) and 
G7V (blue) template. The wavelength limits of the plots are those 
used for the cross--correlation and optimal extraction procedures for 
each object. The region between the dotted lines in the red spectrum of 
V363~Aur was omitted from the cross--correlation and optimal subtraction 
procedures due to the presence of a weak emission line. The average 
eclipse spectra shown (and the corresponding phase ranges listed above) are 
those used to determine the fractional contribution of the secondary 
stars given in Table~\ref{tab:vsini}.}
\label{fig:residual}
\end{figure*}

For AC~Cnc, the spectral types with the lowest $\chi^2$ values are 
G7V and K2V in blue and G9V in red, by no means offering a definitive answer. 
However, the fractional contribution of the secondary star must be less 
than one, ruling out a G type companion. 
By visually inspecting each of the spectra, we settle on 
a spectral type for the secondary star in AC~Cnc of K2 $\pm$ 1V. 
A plot of the AC~Cnc average eclipse 
spectrum, a broadened template spectrum and the residual of the 
optimal subtraction is shown in Fig.~\ref{fig:residual}. The analysis using 
a K2V template results in a $v \sin i$ measurement of 136 km\,s$^{-1}$ in 
blue and 134 km\,s$^{-1}$ 
in red, prompting us to adopt $v \sin i$ = 135 $\pm$ 3 km\,s$^{-1}$. This 
encompasses the $v \sin i$ value for all the G and early--mid K templates 
(except for the blue G5V) within 2$\sigma$. 
The error also reflects all of the other variations noted at the beginning 
of the previous paragraph. 
\scite{schlegel84} conclude that the secondary is a late G or early 
K star, not later than K3. We further limit this to an early K star, most 
likely K2V, agreeing with the studies of \scite{yamasaki83} and 
\scite{zhang87}. The results, however, conflict with the K5 estimate of 
\scite{shugarov81} based on $UBV$ colours. We find that, in eclipse, the 
secondary star in AC~Cnc contributes 85 $\pm$ 5 per cent of the total light 
in the blue and 74 $\pm$ 19 per cent in the red, assuming a K2 $\pm$ 1V 
spectral type. 

For V363~Aur, the G7V 
template yields the lowest $\chi^2$ value in blue, and the G5V proves the 
best in red. Unfortunately, we did not record spectra of the G6V and G7V 
templates in red, so we can only conclude from this analysis and by visual 
inspection that V363~Aur has a secondary of G7 $\pm$ 2V. The average 
$v \sin i$ is 147 $\pm$ 5 km\,s$^{-1}$ in the blue and 
139 $\pm$ 5 km\,s$^{-1}$ in the red. We therefore adopt a compromise value of 
$v \sin i$ = 143 $\pm$ 5 km\,s$^{-1}$, encompassing all $v \sin i$ 
measurements for a G or early--mid K type secondary star. \scite{schlegel86} 
conclude that 
the spectral type is late G, and can be no later than K3, in agreement with 
this study.  We do, however, rule out the estimate of a G0V star by 
\scite{szkody81}.
Using our adopted spectral type of G7 $\pm$ 2V, we find that in 
eclipse the secondary contributes 25 $\pm$ 3 per cent of the light in 
the blue and 45 $\pm$ 8 per cent in the red.

At first glance, the fractional contributions of the secondary 
stars in the two systems during eclipse appear to be inconsistent. The 
K2 $\pm$ 1V secondary star in AC~Cnc contributes a larger fraction 
in blue than red, whereas the (intrinsically bluer) G7 $\pm$ 2V 
secondary in V363~Aur 
contributes a smaller fraction. This can be explained by considering 
the different geometries of the two systems. 
In AC Cnc, almost all of the disc is obscured during eclipse, 
leaving only the redder outer disc uneclipsed. In V363 Aur, however, a 
large portion of the blue inner parts of the disc are still visible 
during eclipse (seen in  Fig.~\ref{fig:roche}), contributing significantly to the 
blue eclipse light. Outside eclipse, we measure the fractional contribution of 
the secondary star in AC~Cnc to be 19 $\pm$ 2 per cent in 
blue and 40 $\pm$ 11 per cent in red. Similar values are found for V363~Aur with 
out-of-eclipse contributions of 19 $\pm$ 4 per cent in blue and 35 $\pm$ 7 
per cent in red.

\begin{table*}
{\protect\small
\caption{$v \sin i$ values (and minimum reduced $\chi^2$) for AC~Cnc and 
V363~Aur cross--correlated with the rotationally--broadened profiles of 
G5--K8 dwarf templates. Degrees of freedom: AC~Cnc, 1143 (blue), 498 
(red); V363~Aur, 580 (blue), 360 (red).
Also shown is the fractional contribution of the secondary star to the 
total light during the eclipse phases, and the position of the strongest peak in the 
skewmaps when cross--correlated with each template using a $\gamma$--velocity 
of 40 km\,s$^{-1}$ for AC~Cnc and 0 km\,s$^{-1}$ for V363~Aur.}
\label{tab:vsini}
\begin{tabular*}{0.98\textwidth}{
cccccr@{$\:\pm\:$}lr@{$\:\pm\:$}lr@{,}lr@{,}l}
(a) AC~Cnc\\
\\
\hline
\vspace{-3mm}\\
\multicolumn{1}{c}{Templates} & 
\multicolumn{1}{c}{min $\chi^{2}$} & 
\multicolumn{1}{c}{$v \sin i$} & 
\multicolumn{1}{c}{min $\chi^{2}$} & 
\multicolumn{1}{c}{$v \sin i$} & 
\multicolumn{2}{c}{Fractional} & 
\multicolumn{2}{c}{Fractional} & 
\multicolumn{2}{c}{($K_x,K_y$) from} & 
\multicolumn{2}{c}{($K_x,K_y$) from} \\
\multicolumn{1}{c}{} & 
\multicolumn{1}{c}{(blue)} & 
\multicolumn{1}{c}{at min $\chi^{2}$} & 
\multicolumn{1}{c}{(red)} & 
\multicolumn{1}{c}{at min $\chi^{2}$} & 
\multicolumn{2}{c}{contribution} & 
\multicolumn{2}{c}{contribution} &
\multicolumn{2}{c}{skew map} &
\multicolumn{2}{c}{skew map} \\
\multicolumn{1}{c}{} & 
\multicolumn{1}{c}{} & 
\multicolumn{1}{c}{(blue)} & 
\multicolumn{1}{c}{} & 
\multicolumn{1}{c}{(red)} & 
\multicolumn{2}{c}{of secondary} & 
\multicolumn{2}{c}{of secondary} &
\multicolumn{2}{c}{(blue)} &
\multicolumn{2}{c}{(red)} \\
\multicolumn{1}{c}{} & 
\multicolumn{1}{c}{} & 
\multicolumn{1}{c}{km\,s$^{-1}$} & 
\multicolumn{1}{c}{} & 
\multicolumn{1}{c}{km\,s$^{-1}$} & 
\multicolumn{2}{c}{(blue)} & 
\multicolumn{2}{c}{(red)} &
\multicolumn{2}{c}{km\,s$^{-1}$} &
\multicolumn{2}{c}{km\,s$^{-1}$} \\
\vspace{-3mm}\\
\hline
\vspace{-3mm}\\
G5V & 1.081 & 142 & 1.106 & 136 & 1.24&0.03 & 1.13&0.10 & (3&187) & (--4&185) \\
G6V & 1.066 & 140 & -- & -- & 1.15&0.03 & \multicolumn{2}{c}{--$\:\:\:$} 
& (3&186) & \multicolumn{2}{c}{--$\:\:\:\:\:\:\:\:\:\:\:\:$} \\
G7V & 1.038 & 140 & -- & -- & 1.18&0.03 & \multicolumn{2}{c}{--$\:\:\:$} 
& (--2&185) & \multicolumn{2}{c}{--$\:\:\:\:\:\:\:\:\:\:\:\:$} \\
G8V & 1.046 & 141 & 1.100 & 137 & 1.20&0.03 & 1.22&0.11 & (--1&186) & (--4&187) \\
G9V & 1.080 & 139 & 1.097 & 136 & 1.19&0.03 & 1.37&0.12 & (3&186) & (--4&186) \\
K0V & 1.043 & 138 & 1.107 & 135 & 1.04&0.02 & 0.99&0.09 & (1&185) & (--4&187) \\
K1V & -- & -- & 1.112 & 134 & \multicolumn{2}{c}{--$\:\:\:$} & 
0.93&0.08 & \multicolumn{2}{c}{--$\:\:\:\:\:\:\:\:\:\:\:\:$} & (--9&186) \\
K2V & 1.038 & 136 & 1.112 & 134 & 0.85&0.02 & 0.74&0.07 & (0&187) 
& (0&185) \\
K3V & 1.127 & 134 & -- & -- & 0.82&0.02 & \multicolumn{2}{c}{--$\:\:\:$} 
& (--2&186) & \multicolumn{2}{c}{--$\:\:\:\:\:\:\:\:\:\:\:\:$} \\
K4V & 1.406 & 138 & 1.137 & 131 & 0.62&0.02 & 0.46&0.04 & (3&189) 
& (--2&185) \\
K5V & 1.288 & 135 & 1.128 & 131 & 0.67&0.02 & 0.56&0.05 & (0&187) 
& (--10&186) \\
K7V & 1.319 & 136 & 1.125 & 130 & 0.65&0.02 & 0.55&0.05 & (4&186) 
& (--3&186) \\
K8V & 1.437 & 136 & 1.138 & 129 & 0.63&0.02 & 0.50&0.05 & (0&188) 
& (--4&184) \\
\hline\\
(b) V363~Aur
\\
\\
\hline
\vspace{-3mm}
\\
G5V & 1.081 & 148 & 1.119 & 139 & 0.25&0.01 & 0.43&0.07 & (--7&184) & (10&185) \\
G6V & 1.081 & 146 & -- & -- & 0.25&0.01 & \multicolumn{2}{c}{--$\:\:\:$} 
& (--8&184) & \multicolumn{2}{c}{--$\:\:\:\:\:\:\:\:\:\:\:\:$} \\
G7V & 1.054 & 148 & -- & -- & 0.27&0.01 & \multicolumn{2}{c}{--$\:\:\:$} 
& (--13&181) & \multicolumn{2}{c}{--$\:\:\:\:\:\:\:\:\:\:\:\:$} \\
G8V & 1.068 & 148 & 1.122 & 138 & 0.26&0.01 & 0.46&0.08 & (--11&184) & (12&184) \\
G9V & 1.098 & 143 & 1.120 & 140 & 0.24&0.01 & 0.52&0.09 & (--4&183) & (13&185) \\
K0V & 1.086 & 143 & 1.126 & 140 & 0.21&0.01 & 0.37&0.07 & (--6&185) & (15&184) \\
K1V & -- & -- & 1.120 & 138 & \multicolumn{2}{c}{--$\:\:\:$} & 0.37&0.06 & 
\multicolumn{2}{c}{--$\:\:\:\:\:\:\:\:\:\:\:\:$} & (7&189) \\
K2V & 1.107 & 142 & 1.136 & 146 & 0.17&0.01 & 0.29&0.06 & (--9&186) & (13&185) \\
K3V & 1.150 & 139 & -- & -- & 0.17&0.01 & \multicolumn{2}{c}{--$\:\:\:$} 
& (--10&184) & \multicolumn{2}{c}{--$\:\:\:\:\:\:\:\:\:\:\:\:$} \\
K4V & 1.294 & 134 & 1.140 & 146 & 0.10&0.01 & 0.19&0.04 & (0&191) & (9&186) \\
K5V & 1.234 & 134 & 1.132 & 141 & 0.11&0.01 & 0.24&0.04 & (5&188) & (8&188) \\
K7V & 1.252 & 132 & 1.134 & 145 & 0.11&0.01 & 0.23&0.04 & (0&191) & (12&187) \\
K8V & 1.297 & 134 & 1.138 & 148 & 0.10&0.01 & 0.22&0.04 & (--4&188) & (8&187) \\
\vspace{-3mm}\\
\hline
\end{tabular*}
}
\end{table*}

\subsection{The $K_R$ Correction}
\label{sec:kcor}

The irradiation of the secondary stars in CVs by the emission regions around 
the white dwarf and the bright spot has been shown to influence the measured 
$K_R$ (e.g. \pcite{wade88} and \pcite{watson00}). For example, if absorption 
lines are quenched on the irradiated side of the secondary, the 
centre of light will be shifted towards the back of the star. The measured 
$K_R$ will then be larger than the true (dynamical) value. 

We must now determine whether the secondary stars in AC~Cnc and V363~Aur 
are irradiated, which can be observationally tested in two ways. 
Firstly, the rotationally broadened line profile would be distorted if 
there was a non--uniform absorption distribution across the surface of the 
secondary star (\pcite{davey92}). This would result in a non--sinusoidal radial 
velocity curve. Secondly, one would expect a depletion of absorption line flux 
from the secondary star 
at phase 0.5, where the quenched inner--hemisphere is pointed towards 
the observer (e.g. \pcite{friend90a}). We applied these tests to the AC~Cnc and 
V363~Aur data.

\begin{figure*}
\begin{center}
\begin{tabular}{c}
\multicolumn{1}{l}{(a) AC~Cnc}\\
\includegraphics[width=10.2cm,angle=-90]{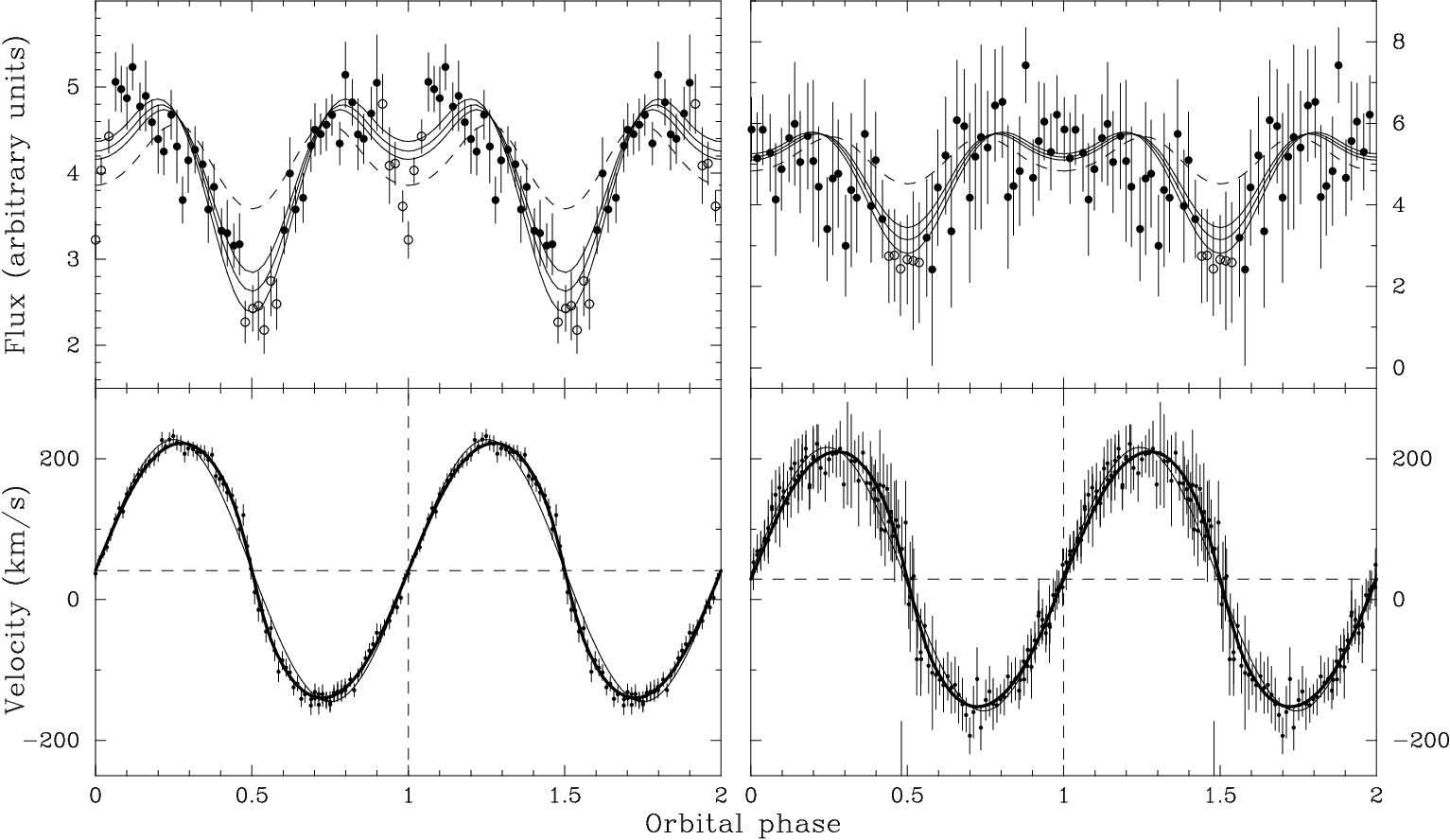}\\
\multicolumn{1}{l}{(b) V363~Aur}\\
\includegraphics[width=10.2cm,angle=-90]{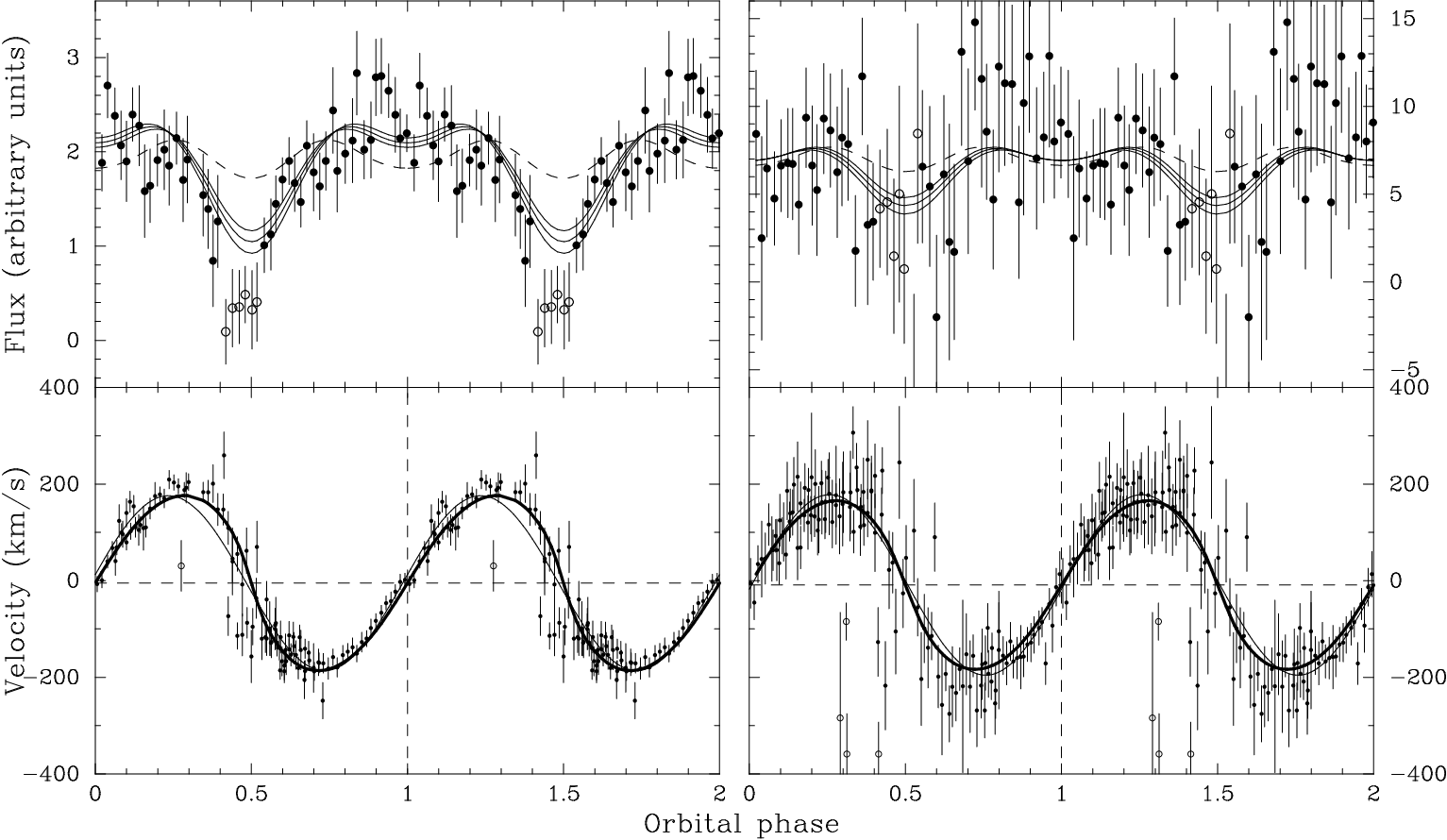}\\
\end{tabular}
\end{center}
\caption{
Upper panel: Secondary star lightcurves with model fits 
(solid lines). For AC~Cnc, model fits are shown for Roche lobes with 7, 8 
and 9 slices removed in both blue (left) and red (right). For V363~Aur, 
model fits are shown for Roche lobes with 9, 10 and 11 slices removed in 
blue (left) and 7, 8 and 9 in red (right). The lower the 
line, the more slices removed. The dashed line in all plots represents a model 
where 0 slices have been removed. The data have been phase--binned into 
50 bins to increase signal--to--noise. 
Lower panel: Measured secondary star radial--velocity curve with a sinusoidal fit 
(thin solid line) and the best--fitting model fit (thick solid line). 
In all panels, the open circles indicate points that have been omitted from 
the fits and all data have been folded to show 2 orbital phases.}
\label{fig:irrplot}
\end{figure*}

The secondary star radial velocity curves were produced by cross--correlating 
the CV spectra with the best--fitting smeared and broadened template spectra as 
described in Section~\ref{sec:secondary}. This time the cross--correlation 
peaks were plotted against phase to produce the radial velocity curves shown 
in the lower panels of Fig.~\ref{fig:irrplot}. 
The radial velocity curves of both AC~Cnc and V363~Aur are clearly eccentric 
in comparison to the sinusoidal fits represented by the thin solid lines. 

The variation of secondary star absorption line flux with phase for AC~Cnc and 
V363~Aur is shown in the top panels of 
Fig.~\ref{fig:irrplot}. These lightcurves were produced by 
optimally subtracting the smeared and rotationally broadened 
best--fitting template from the individual CV spectra (with the secondary 
radial velocity shifted out) as described in Section~\ref{sec:rotational}. 
This time, however, the 
spectra were continuum subtracted rather than normalised to ensure that the 
measurements were not affected by a fluctuating disc brightness. The constants 
produced by the optimal subtraction are secondary star absorption line fluxes, 
correct relative to each other, but not in an absolute sense. 
The dashed lines super--imposed on the lightcurves represent the variation of 
flux with phase for a Roche lobe with a uniform absorption distribution. The 
sinusoidal nature is the result of the changing projected area of the 
Roche lobe through the orbit. The lightcurves of AC~Cnc and 
V363~Aur exhibit a drop in flux at phase 0.5 in comparison with the 
uniform Roche lobe. 

These two pieces of evidence, as well as the observed Balmer emission from the inner 
hemisphere of the secondary stars seen in Figs.~\ref{fig:trailed} 
and ~\ref{fig:doppler}, and the weakening of the CCFs around phase 0.5 seen in 
Fig.~\ref{fig:skewmaps} suggest that the secondary stars in AC~Cnc and 
V363~Aur are indeed irradiated and we must correct the $K_R$ values accordingly.

It is possible to correct $K_R$ for the effects of irradiation by 
modelling the secondary star absorption line flux distribution. In our model, 
we divided the secondary Roche lobe into 40 vertical slices of equal width. 
We then produced a series of model lightcurves, varying the numbers of 
slices omitted from the inner hemisphere of the secondary 
which contribute to the total flux. 
The model lightcurves were then scaled to match the observed data, and the 
best--fitting model found by measuring the $\chi^2$ between the two.
In all models, we used a gravity--darkening parameter $\beta = 0.08$ and 
limb--darkening coefficient $u = 0.5$ (e.g. \pcite{watson00}). 
There is evidence for a secondary eclipse in both systems, so we have 
omitted points around phase 0.5 from the fits. (We tried a model which included 
an accretion disc to reproduce the secondary eclipse, but the results were 
exactly the same as omitting the points.) 
Once the best--fitting lightcurve was found, 
we produced fake CV spectra from the models, which were 
cross-correlated with a fake template star to produce a synthetic radial 
velocity curve. 
In the first instance, the synthetic curve mimicked the non-sinusoidal 
nature of the observed data, but with a larger semi--amplitude. This was 
expected, as the model input parameters used the uncorrected 
$K_R$ derived in Section~\ref{sec:params}. We then lowered $K_R$ and repeated 
the process, until the 
semi--amplitude of the model and observed radial velocity curves were in 
agreement, each time checking the lightcurve models for goodness of fit. 
The resulting $K_R$ was 
then adopted as the real (or dynamical) $K_R$ value. 

For AC~Cnc, the best--fitting model lightcurve was produced by omitting 
8 slices when fitting both the blue and red data. The model lightcurves omitting 
7, 8 and 9 slices are shown by the solid lines in Fig.~\ref{fig:irrplot}. 
Our final model, which has an input $K_R$ of 176 km\,s$^{-1}$, produces the 
radial velocity curve shown as the thick solid line in Fig.~\ref{fig:irrplot}.
There is excellent agreement between this and the observed data. 
The best--fitting lightcurve models for V363~Aur have 10 slices omitted 
in blue and 8 slices in red. Model lightcurves omitting 9, 10 and 11 
slices in blue and 7, 8 and 9 slices in red are once again shown as solid 
lines in Fig.~\ref{fig:irrplot}. The red data are very noisy, so we 
used the blue data to obtain a corrected $K_R$ of 168 km\,s$^{-1}$. 
The model with this input value has the radial velocity curve plotted as the 
thick solid line in Fig.~\ref{fig:irrplot}.
It should be noted that if gravity--darkening and limb--darkening are 
neglected, the best fit lightcurves remain the same in all cases, and 
produce $K_R$ values which are $\sim$ 3 km\,s$^{-1}$ lower.

The points around primary eclipse in the blue lightcurve of AC~Cnc were 
also omitted from the above fit, as they show a very sharp decrease 
in flux. Although some of this can be attributed to the reduced projected 
area of the secondary at phase 0, the feature is too sharp and too deep for 
this to be the only explanation. 
It is likely that this feature is an artefact of the slit--loss 
correction procedure, where $B$--band photometry has been used to correct 
secondary features which actually lie closer to the $V$--band. Because the 
$B$--band eclipse is deeper than the $V$--band eclipse, we get a residual 
sharp dip in the secondary star flux at phase 0.
There is no corresponding dip in the 
red data, as the photometry and spectroscopy wavelengths 
are closely matched. There is no corresponding dip in flux at 
phase 0 in the blue V363~Aur data, even though both objects have been 
reduced in the same way. We believe this is because the $V$--band eclipse 
depth is approximately the same as the $B$--band eclipse depth in V363~Aur. 

In summary, we correct the $K_R$ of AC~Cnc from 186 km\,s$^{-1}$ 
to 176 km\,s$^{-1}$ and the $K_R$ of V363~Aur from 
184 km\,s$^{-1}$ to 168 km\,s$^{-1}$.

\subsection{The distances to AC~Cnc and V363~Aur}
\label{sec:distance}

By finding the apparent magnitude of the secondary star from its 
contribution to the total light during eclipse, and estimating its 
absolute magnitude, we can calculate the distance ($d$) to each system 
using the equation: 

\begin{equation}
\label{eqn:distance}
{5 \log(d/10) = m_V - M_V - d\;A_V/1000}
\end{equation}

\noindent{where $A_V$ is the visual interstellar extinction in 
magnitudes per kpc.}

There are a number of ways of estimating the absolute magnitude of the 
secondary star, assuming it is on the main sequence (e.g. 
\pcite{patterson84}; \pcite{warner95b}; \pcite{gray92}). 
The distance estimates given below take into account all of 
these techniques. 

Another method of finding the distance is to determine the angular diameter 
of the secondary star from the observed flux and a surface brightness 
calibration that we derive from the Barnes--Evans relation (\pcite{barnes76}),

\begin{equation}
\label{eqn:barnesevans}
{F_v = 4.2207 - 0.1V_0 - 0.5\log\phi = 3.977 - 0.429(V-R)_0}
\end{equation}

\noindent{where $V_0$ and $(V-R)_0$ are the unreddened $V$ magnitude and 
$(V-R)$ colour of the secondary star, and $\phi$ is the stellar angular 
diameter in arc milliseconds.}

\subsubsection{AC~Cnc}
\label{sec:accnc_dist}

During the eclipse phases given in Section~\ref{sec:rotational}, the average 
apparent magnitude of AC~Cnc is $R$ = 14.7 $\pm$ 0.1, of which the 
secondary contributes 74 $\pm$ 19 per cent, and 
$B$ = 15.9 $\pm$ 0.1, of which the secondary 
star contributes 85 $\pm$ 5 per cent. The apparent magnitude of the secondary 
is therefore $R$ = 15.0 $\pm$ 0.3 and $B$ = 16.1 $\pm$ 0.1. 
Using typical $B-V$ and $V-R$ values for an early K star from \scite{gray92}, 
we arrive at an apparent $V$ magnitude of 15.5 $\pm$ 0.3. 
We adopt an 
absolute magnitude of $M_V = +6.8 \pm 0.5$ as an average of the various 
methods referenced in the previous paragraph. Assuming zero interstellar 
extinction (\pcite{ladous91}), the distance to AC~Cnc is 550 $\pm$ 150 pc.

Using the Barnes--Evans relation with a $(V-R)_0$ value typical of an 
early K star (0.74 $\pm$ 0.10; \pcite{gray92}), the $V_0$ value of 
15.5 $\pm$ 0.3 found above and the radius of the secondary star derived in 
Section~\ref{sec:params}, we obtain a distance of 750 $\pm$ 250 pc.

Published estimates of the distance to AC~Cnc agree with our calculated 
values. \scite{shugarov81} calculates a distance of 480 pc assuming a K5 
secondary star. \scite{patterson84} derives a value of 400 pc by combining an 
$M_V$--H$\beta$ equivalent--width relationship, properties of the secondary 
and the continuum shape of the spectrum. \scite{warner87} calculates a 
distance of 800 pc using the secondary star characteristics from \scite{yamasaki83} 
and \scite{zhang87} concludes $d$ = 500 $\pm$ 100 pc, again estimating the secondary 
star properties.

\subsubsection{V363~Aur}
\label{sec:v363aur_dist}

The average apparent magnitude of V363~Aur during the eclipse phases 
quoted in Section~\ref{sec:rotational} is $B$ = 14.8 $\pm$ 0.1 and 
$R$ = 13.9 $\pm$ 0.1, of which the G7 $\pm$ 2V secondary star contributes 
25 $\pm$ 3 per cent and 45 $\pm$ 8 per cent, respectively. 
This corresponds to an apparent magnitude of the secondary 
star of $B$ = 16.3 $\pm$ 0.2, $R$ = 14.8 $\pm$ 0.2 and, assuming typical 
$B-V$ and $V-R$ values for a G7 $\pm$ 2V star (\pcite{gray92}), we calculate 
$V$ = 15.4 $\pm$ 0.2. There is evidence of interstellar absorption in the 
average spectrum (Fig.~\ref{fig:average}), 
suggesting that it is important to take extinction into account. 
\scite{szkody81} give a value of $E(B-V)$ = 0.3, although the 
UV spectrum used was underexposed and noisy. \scite{ladous91}, however, 
finds no extinction to V363~Aur, a result contradicted by measurements of 
\scite{rutten92b}, who measure $E(B-V)$ = 0.1. We adopt this 
value, which results in $A_V = 0.32$ (\pcite{scheffler82}). 
We use $M_V = +6.0 \pm 0.5$ as an average absolute magnitude of the 
G7 $\pm$ 2V star secondary star in V363~Aur, which results in a distance of 
700 $\pm$ 250 pc.

Using the Barnes--Evans relation with a $(V-R)_0$ value typical of a 
mid--late G star (0.56 $\pm$ 0.1; \pcite{gray92}), a $V_0$ value of 
15.1 $\pm$ 0.2 assuming extinction to be 0.3 mag and the radius of the 
secondary star derived in Section~\ref{sec:params}, we obtain a distance of 
1000 $\pm$ 250 pc.

\scite{szkody81} estimate a distance to V363~Aur of 900 pc assuming a 
G0 dwarf secondary. \scite{patterson84} 
derives an uncertain value of 1000 pc from the $M_V$--H$\beta$ 
equivalent--width relationship, interstellar absorption and the continuum 
shape of the spectrum. In his compilation of distances, 
\scite{berriman87} gives a distance of 1100 pc from disc properties, 
900--1300 pc from red star spectrophotometry and $>$450pc from the infrared 
properties of the secondary. \scite{rutten92b} calculate a distance 
of 530 pc from a black body fit to the spectrum of the central part of the 
disc and 720 pc using a value for the fractional contribution of the secondary.

\subsection{System Parameters}
\label{sec:params}

Using the $K_R$ and $v \sin i$ values found in Sections~\ref{sec:rotational} 
and~\ref{sec:kcor} in conjunction with the period determined in 
Section~\ref{sec:ephem} and a measurement of the 
eclipse full--width at half depth ($\Delta\phi_{1/2}$), we can calculate 
accurate system parameters for AC~Cnc and V363~Aur.

In order to determine $\Delta\phi_{1/2}$, we estimated the flux out of 
eclipse (the principal source of error) and at eclipse minimum, 
and then measured the full--width of the eclipse half-way between these 
points. The eclipse full--width at half-depth was measured to 
be $\Delta\phi_{1/2}$ = 0.096 $\pm$ 0.002 for AC~Cnc and 0.063 $\pm$ 0.002 
for V363~Aur from the $B$ and $R$--band lightcurves in Fig.~\ref{fig:lc}.

We have opted for a Monte Carlo approach similar to \scite{horne93} to 
calculate the system parameters and their errors. For a given set 
of $K_R$, $v \sin i$, $\Delta\phi_{1/2}$ and $P$, the other system parameters 
are calculated as follows.

$R_2/a$ can be estimated because we know that the secondary star fills its 
Roche lobe (as there is an accretion disc present and hence mass transfer). 
$R_2$ is the equatorial radius of the secondary star and $a$ is the binary 
separation. We used Eggleton's formula (\pcite{eggleton83}) which gives the 
volume-equivalent radius of the Roche lobe to better than 1 per cent, which 
is close to the equatorial radius of the secondary star as seen during 
eclipse,
\begin{equation}
{{R_2} \over a} = {{0.49q^{2/3}} \over {{0.6q^{2/3} + \ln{(1+q^{1/3})}}}}. 
\label{eqn:eggleton}
\end{equation}
The secondary star rotates synchronously with the orbital motion, so 
we can combine $K_R$ and $v \sin i$, to get
\begin{equation}
{{R_2} \over a}{(1 + q)} = {{v \sin i} \over {K_R}}.
\label{eqn:synch}
\end{equation}
By considering the geometry of a point eclipse by a spherical body 
(e.g. \pcite{dhillon91}), the radius of the secondary can be shown to be
\begin{equation}
\biggl({{R_2} \over a}\biggr)^2 = \sin^2\pi\Delta\phi_{1/2}+
\cos^2\pi\Delta\phi_{1/2}\cos^2i,
\label{eqn:inclin}
\end{equation}
which, using the value of $R_2/a$ obtained using equations~\ref{eqn:eggleton} 
and~\ref{eqn:synch}, allows us to calculate the inclination ($i$) of the system. 
The geometry of a disc eclipse can be approximated to a point eclipse if the 
light distribution around the white dwarf is axi--symmetric (e.g. 
\pcite{dhillon90}). This approximation is justified given the symmetry of the 
primary eclipses in the photometry light curves (Figure~\ref{fig:lc}). 
Kepler's Third Law gives us
\begin{equation}
{{K_R^3P_{orb}}\over{2\pi G}}={{M_1\sin^3i}\over{(1+q)}^2},
\end{equation}
which, with the values of $q$ and $i$ calculated using equations 
~\ref{eqn:eggleton},~\ref{eqn:synch} and~\ref{eqn:inclin}, gives the mass 
of the primary star. The mass of the secondary star can then be obtained using 
\begin{equation}
{M_2} = {q{M_1}}.
\label{eqn:qratio}
\end{equation}
The radius of the secondary star is obtained from the equation 
\begin{equation}
{{v \sin i} \over {R_2}} = {{2\pi \sin i} \over P},
\label{eqn:secradius}
\end{equation}
(e.g. \pcite{warner95a}) and the separation of the components, $a$, 
is calculated from equations 
~\ref{eqn:synch} and ~\ref{eqn:secradius} with $q$ and $i$ now known.

\begin{figure*}
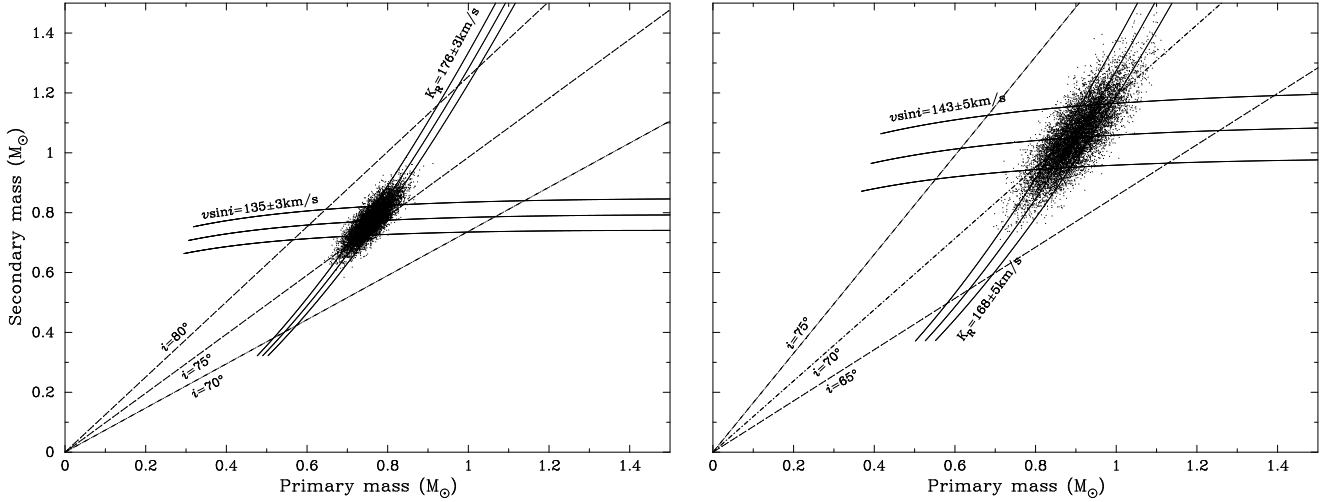

\begin{center}
\begin{tabular}{lll}
\includegraphics[width=6.6cm,angle=-90]{accnc_kcorrected.ps}
\hspace{3mm}
\includegraphics[width=6.6cm,angle=-90]{v363aur_kcorrected.ps}
\end{tabular}
\end{center}
\caption{\protect\small
Monte Carlo determination of system paramters for AC~Cnc (left) 
and V363~Aur (right). Each dot represents an $M_1,M_2$ pair; the solid 
curves satisfy the $v \sin i$ and corrected $K_R$ constraints, and the dashed 
lines mark lines of constant inclination.}
\label{fig:montecarlo}
\end{figure*}

The Monte Carlo simulation takes 10\,000 values of $K_R$, $v \sin i$, and 
$\Delta\phi_{1/2}$ (the error on the period is deemed to be negligible in 
comparison to the errors on $K_R$, $v \sin i$, and $\Delta\phi_{1/2}$), 
treating each as being normally distributed about their 
measured values with standard deviations equal to the errors on the 
measurements. We then calculate the masses of the components, the inclination 
of the system, the radius of the secondary star, and the separation of the 
components, as outlined above, omitting 
($K_R$, $v \sin i$, $\Delta\phi_{1/2}$) 
triplets which are inconsistent with $\sin i \leq1$. Each accepted 
$M_1,M_2$ pair is then plotted as a point in Figure~\ref{fig:montecarlo}, 
and the masses and their errors are computed from 
the mean and standard deviation of the distribution of these pairs. 

In the case of AC~Cnc, we find that 
$M_1 = 0.76 \pm 0.03M_\odot$ and $M_2 = 0.77 \pm 0.05M_\odot$; for V363~Aur 
we find $M_1 = 0.90 \pm 0.06M_\odot$ and $M_2 = 1.06 \pm 0.11M_\odot$.
The values of all the system parameters deduced from the Monte Carlo 
computation are listed in Table~\ref{tab:params}, including $K_R$--corrected 
and non $K_R$--corrected values for comparison.

We computed the radius of the accretion discs in AC~Cnc and V363~Aur using 
the geometric method outlined in \scite{dhillon91}. The phase half-width of 
eclipse at maximum intensity was found in Section~\ref{sec:light} to be 
$\Delta\phi = 0.09 \pm 0.01$ 
for AC~Cnc and $\Delta\phi = 0.078 \pm 0.005$ for V363~Aur. Combining 
$\Delta\phi$ with $q$ and $i$ derived above produces an accretion disc 
radius ($R_D$) in terms of the volume radius of the primary's Roche lobe 
($R_1$). We find accretion disc radii of $R_D/R_1 = 0.61 \pm 0.14$ and 
$R_D/R_1 = 0.71 \pm 0.17$ for AC~Cnc and V363~Aur, respectively, which 
are lower than those quoted by \scite{harrop96}, but consistent within the errors; 
AC~Cnc: $0.70 \le R_D/R_1 \le 0.86$, V363~Aur: $R_D/R_1 \ge 0.92$.

The empirical relation obtained by \scite{smith98} between mass and radius 
for the secondary stars in CVs is given by,
\begin{equation}
\label{eq:smith}
{R \over R_\odot} = 
(0.93\pm0.09)\Biggl({M \over M_\odot}\Biggr)+(0.06\pm0.03).
\end{equation}
This predicts that if the secondary stars in AC~Cnc and V363~Aur are on 
the main-sequence, they should have radii of 0.78$R_\odot$ and 1.05$R_\odot$, 
respectively. These values agree with our measured values of 
0.83 $\pm$ 0.03 $R_\odot$ and 0.97 $\pm$ 0.04 $R_\odot$ to within the 
errors. \scite{gray92} gives $M = 0.76M_\odot$ and $R = 0.75R_\odot$ for a 
K2 dwarf and $M = 0.98M_\odot$ and $R = 0.96R_\odot$ for a G5 dwarf, 
also in agreement with our measured values. We conclude that the secondaries 
in AC~Cnc and V363~Aur are similar to main--sequence stars.

\begin{table*}
{\protect\small
\caption[System parameters]{System parameters for AC~Cnc and V363~Aur. The 
Monte Carlo results for corrected and uncorrected $K_R$ values are 
shown for comparison. The radial velocity of the white dwarf ($K_W$) has 
also been calculated from the secondary star parameters. Distances are 
quoted using both techniques described in the text; by 
estimating the secondary star magnitude ($d^{sec}$) and using the Barnes--
Evans relation ($d^{B-E}$).}
\label{tab:params}
\begin{tabular*}{155mm}
{lr@{$\:\pm\:$}lr@{$\:\pm\:$}lr@{$\:\pm\:$}lr@{$\:\pm\:$}lr@{$\:\pm\:$}lr@{$\:\pm\:$}l} 
\\
\hline
\vspace{-3mm}\\
\multicolumn{1}{l}{Parameter} &
\multicolumn{6}{c}{AC Cnc} & 
\multicolumn{6}{c}{V363 Aur} \\
\multicolumn{1}{l}{} &
\multicolumn{12}{c}
{---------------------------------------------------------------------------------------------------------------------------------}\\  
\multicolumn{1}{l}{} &
\multicolumn{2}{c}{Measured} & 
\multicolumn{2}{c}{Monte Carlo} & 
\multicolumn{2}{c}{$K_R$--corrected} & 
\multicolumn{2}{c}{Measured} & 
\multicolumn{2}{c}{Monte Carlo} & 
\multicolumn{2}{c}{$K_R$--corrected} \\
\multicolumn{1}{l}{} &
\multicolumn{2}{c}{Value} & 
\multicolumn{2}{c}{Value} &
\multicolumn{2}{c}{Value} & 
\multicolumn{2}{c}{Value} & 
\multicolumn{2}{c}{Value} & 
\multicolumn{2}{c}{Value}  \\
\vspace{-3mm}\\
\hline
\vspace{-3mm}\\
$P_{orb}$ (d) & \multicolumn{2}{c}{0.30047747} & \multicolumn{2}{c}{} & 
\multicolumn{2}{c}{0.30047747} & 
\multicolumn{2}{c}{0.32124187} & \multicolumn{2}{c}{} & 
\multicolumn{2}{c}{0.32124187} \\
$K_R$ (km\,s$^{-1}$) & 186&3 & 186&3 & 176&3 
& 184&5 & 184&5 & 168&5 \\
$v \sin i$ (km\,s$^{-1}$) & 135&3 & 135&3 & 135&3 
& 143&5 & 143&5 & 143&5 \\
$\Delta\phi_{1/2}$ & 0.096&0.003 & 0.096&0.003 & 0.096&0.003 
& 0.063&0.002 & 0.063&0.002 & 0.063&0.002 \\
$q$ & \multicolumn{2}{c}{} & 0.94&0.04 & 1.02&0.04 
& \multicolumn{2}{c}{} & 1.04&0.06 & 1.17&0.07 \\
$i^\circ$ & \multicolumn{2}{c}{} & 76.3&0.8 & 75.6&0.7 
& \multicolumn{2}{c}{} & 70.5&0.4 & 69.7&0.4 \\
$K_W$ (km\,s$^{-1}$) & \multicolumn{2}{c}{} & 175&6 & 179&6 
& \multicolumn{2}{c}{} & 190&9 & 196&9 \\
$M_1/M_\odot$ & \multicolumn{2}{c}{} & 0.82&0.04 & 0.76&0.03 
& \multicolumn{2}{c}{} & 1.03&0.07 & 0.90&0.06 \\
$M_2/M_\odot$ & \multicolumn{2}{c}{} & 0.78&0.05 & 0.77&0.05 
& \multicolumn{2}{c}{} & 1.06&0.11 & 1.06&0.11 \\
$R_2/R_\odot$ & \multicolumn{2}{c}{} & 0.82&0.02 & 0.83&0.02 
& \multicolumn{2}{c}{} & 0.96&0.04 & 0.97&0.04 \\
$a/R_\odot$ & \multicolumn{2}{c}{} & 2.21&0.04 & 2.18&0.04 
& \multicolumn{2}{c}{} & 2.52&0.07 & 2.47&0.07 \\
$d ^{sec}$ (pc) & 550&150 & \multicolumn{2}{c}{} & \multicolumn{2}{c}{} & 700&250 
& \multicolumn{2}{c}{} & \multicolumn{2}{c}{} \\
$d ^{B-E}$ (pc) & 750&250 & \multicolumn{2}{c}{} & \multicolumn{2}{c}{} & 1000&250 
& \multicolumn{2}{c}{} & \multicolumn{2}{c}{} \\
Spectral type & K2&1 V & \multicolumn{2}{c}{} & 
\multicolumn{2}{c}{} & G7&2 V
& \multicolumn{2}{c}{} &  \multicolumn{2}{c}{} \\
of secondary & \multicolumn{2}{c}{} & \multicolumn{2}{c}{} 
& \multicolumn{2}{c}{} & \multicolumn{2}{c}{} 
& \multicolumn{2}{c}{} & \multicolumn{2}{c}{} \\
$\Delta\phi$ & 0.09&0.01 & \multicolumn{2}{c}{} & \multicolumn{2}{c}{}
& 0.078&0.005 & \multicolumn{2}{c}{} & \multicolumn{2}{c}{} \\
$R_D/R_1$ & \multicolumn{2}{c}{} & \multicolumn{2}{c}{} 
& 0.61&0.14 & \multicolumn{2}{c}{} 
& \multicolumn{2}{c}{} & 0.71&0.17 \\
\hline\\
\end{tabular*}
}
\end{table*}

\section{Discussion}
\label{sec:discussion}

\subsection{Are AC~Cnc and V363~Aur SW~Sex stars?}
\label{sec:swsex}

The SW~Sex stars are a sub--class of NLs which have peculiar spectral 
properties -- see \pcite{hellier00} and references therein for a summary of the 
current models. Most SW~Sex stars share the following properties 
(e.g. \pcite{hoard03}): 

\begin{enumerate}

\item{They are usually high--inclination, eclipsing systems.}

\item{Their spectra exhibit single--peaked emission lines rather 
than the double--peaked lines expected from near edge--on discs.}

\item{The Balmer and HeI emission lines usually contain transient absorption 
cores, especially around photometric phase 0.5.}

\item{They have high levels of excitation, with HeII $\lambda$4686\AA$\:$ 
emission often comparable in strength to H$\beta$.}

\item{The low--excitation lines (Balmer and HeI) exhibit shallow or absent 
eclipses.}

\item{High velocity S--waves are often seen in the trailed spectra.}

\item{The emission--line radial velocity curves show large phase shifts 
between spectroscopic conjunction and photometric mid--eclipse.}

\end{enumerate}

AC~Cnc and V363~Aur show many of the features described above. 
V363~Aur shows all of the 
features, so must be considered a definite SW~Sex star. AC~Cnc does 
show high excitation features, but not to the extent of V363~Aur.
Transient absorption features are seen in low--excitation lines of both 
systems. An interesting absorption feature also occurs around phase 0
in V363~Aur, which deserves further discussion.

\begin{figure*}
\begin{center}
\includegraphics[width=4cm,angle=-90]{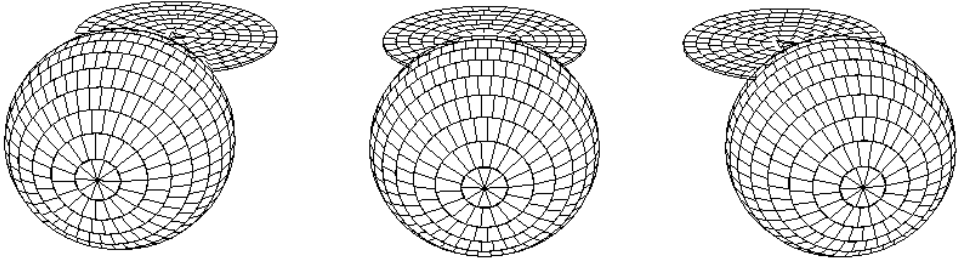}
\end{center}
\caption{\protect\small
A model of V363~Aur using the $K_R$--corrected system parameters listed in 
Table~\ref{tab:params}. During primary eclipse, 
the secondary star grazes the absorption region at the centre of the disc, 
resulting in the rapid increase and then rapid decrease in Balmer flux seen 
in the trailed spectra at phase 0. The left--hand diagram shows the system at 
phase 0.97, the centre diagram at phase 1.00 and the right--hand diagram at 
phase 1.03.}
\label{fig:roche}
\end{figure*}

The low--excitation emission lines of V363~Aur show the broad primary 
eclipse expected from a gradual obscuration of the accretion disc by the 
secondary (see Fig.\ref{fig:lc}). Around phase 0, however, there is a sharp 
increase in Balmer line 
emission followed by a rapid decrease. The feature comes and goes too 
rapidly to be attributable to a permanent emission source on the back of the 
secondary star. 
We suggest that this 
increase in flux can be explained by the eclipse of an optically thick 
region at the centre of the disc. The eclipse of this region, which would 
have an absorption line spectrum, would result in an overall increase in the 
line flux at the phases observed. Fig.~\ref{fig:trailedzoom} is the trailed 
spectrum of H$\beta$ magnified to show the effect more clearly.  
The blue wing of the line increases in flux first, 
in agreement with an eclipse of the centre of the disc. Fig.~\ref{fig:roche} is 
a reconstruction of V363 Aur using the $K_R$--corrected system parameters 
listed in Fig.~\ref{tab:params}. The left--hand diagram shows the system just 
before the increase in flux at phase 0.97, where 
the centre of the disc is just about to be eclipsed by the secondary. In 
the central diagram, the centre of the disc is eclipsed, corresponding to 
the brief rise in line flux. The right--hand diagram shows the system at 
phase 1.03, when the centre of the 
disc comes out of eclipse, and the line flux drops again. The short timescale 
of this feature is explained by the grazing nature of the eclipse of the 
optically thick region at the centre of the disc. This model could also 
account for the behaviour of other SW~Sex stars where the Balmer lines 
show only a shallow eclipse compared to the continuum. 
\scite{dhillon97b} and \scite{groot01} suggest a similar explanation 
for SW~Sex. The latter authors invoked an optically thick absorption--line 
source coincident with the bright spot, similar to that of a late B or early 
A--type star. 

\begin{figure}
\begin{center}
\includegraphics[width=6cm,angle=-90]{trailedzoom.ps}
\end{center}
\caption{\protect\small
The V363~Aur H$\beta$ trailed spectrum, magnified to show the phases 
0.8--1.2. The emission feature is clearly seen around phase 0, moving from 
blue to red.}
\label{fig:trailedzoom}
\end{figure}

\subsection{Mass Transfer Stability}
\label{sec:massratio}

\begin{figure}
\begin{center}
\includegraphics[width=8cm,angle=0]{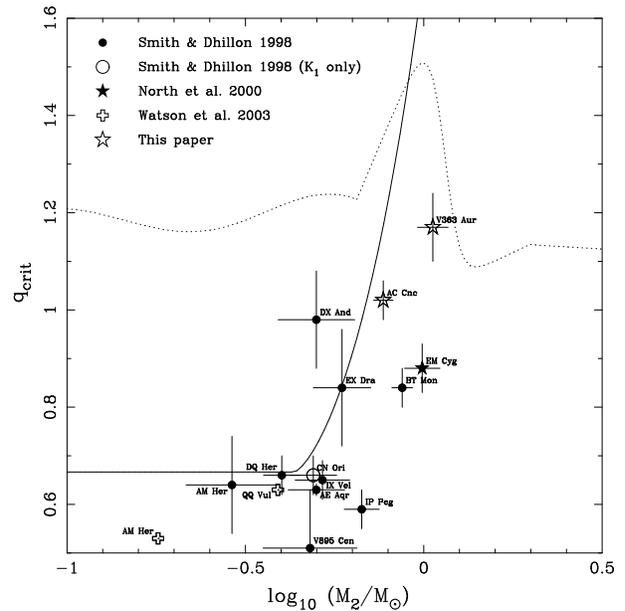}
\caption{\protect\small
Critical mass ratios for mass transfer stability. The dotted line represents 
the condition for thermal instability; the solid line represents the 
condition for dynamical instability (Politano 1996). 
Both curves assume the star is initially in thermal equilibrium. 
Mass ratios and secondary masses from the compilation of 
\protect\scite{smith98}, \protect\scite{north00} and \protect\scite{watson03} 
are overplotted. The mass ratios and secondary star masses of AC~Cnc 
and V363~Aur determined in this paper are also plotted.}
\label{fig:qplot}
\end{center}
\end{figure}

The mass ratio ($q = M_2/M_1$) of a CV is of great significance, as it 
governs the properties of mass transfer from the secondary to the white 
dwarf primary. This in turn governs the evolution and behaviour 
of the system. 

The secondary
star has two timescales on which it responds to mass loss.
Firstly, the star returns to hydrostatic equilibrium on a dynamical
timescale, which is the  sound--crossing time of the region
affected. Secondly, on a
longer timescale, the star settles into a new thermal equilibrium
configuration. The second timescale a star responds on is therefore
the thermal, or Kelvin-Helmholtz, timescale.

The two timescales upon which the secondary responds to mass loss leads to
two types of mass transfer instability. If, upon mass loss, the dynamical
response of the secondary is to expand relative to the Roche lobe, mass
transfer is dynamically unstable and mass transfer proceeds on the
dynamical timescale.
\scite{politano96} made an analytic fit 
to the models of \scite{hjellming89} to give the limit of dynamically stable 
mass loss, plotted as the solid line in Fig.~\ref{fig:qplot}. Dynamically 
stable mass transfer can occur if the CV lies below this line. 
This limit is important for low mass secondary stars ($M_2 < 0.5M_\odot$), as 
they have significant convective envelopes that tend to expand adiabatically 
in response to mass loss (\pcite{dekool92}). 
 
Thermally unstable mass transfer is possible if the dynamic response
of the star to mass loss is to shrink relative to its Roche lobe
(i.e. mass transfer is {\em dynamically} stable). 
This occurs at high donor masses ($M_2 > 0.8M_\odot$) when the 
star has a negligible convective envelope and its adiabatic response 
to mass loss is to shrink (e.g. \pcite{dekool92}; \pcite{politano96}). 
Mass transfer then
initially breaks contact and the star begins to settle into its new thermal
equilibrium configuration. If the star's thermal equilibrium radius is
now bigger than the Roche lobe, mass transfer is again unstable, but
proceeds on the slower, thermal timescale. 
The limit of thermally stable mass transfer can be 
found by differentiating the main--sequence mass--radius relationship 
given in \scite{politano96}.
Thermally stable mass transfer can occur if the CV appears below the dotted 
line plotted in Fig.~\ref{fig:qplot}. 

Most CVs on
the plot fall below both curves, implying that mass transfer is
dynamically and thermally stable, as expected (the mass transfer rates
observed in CVs are too low for unstable mass transfer to be
occurring). An exception is DX And, which appears to be dynamically
unstable. This is, of course, not possible; a system undergoing
dynamical mass transfer would rapidly form a common envelope. A
solution is found in the fact that DX And has an evolved secondary
star (\pcite{drew93}); the dynamical and thermal solutions plotted on 
Fig.~\ref{fig:qplot} are for ZAMS stars, so cannot be applied to 
evolved stars. 

The previous study of the masses of AC~Cnc by \scite{schlegel84} found a 
$q$ value of 1.24 $\pm$ 0.08, close to the limit 
for thermally stable mass transfer. In this study, we find a lower value 
of $q$ = 1.02 $\pm$ 0.04, placing it well within the theoretical limit. 
In the case of V363~Aur, we find a 
higher $q$ value of 1.17 $\pm$ 0.07 than the value of $q$ = 0.90 $\pm$ 0.10 
quoted by \scite{schlegel86}, but it is still within the stability limit. 
The mass ratios found in this paper therefore place AC~Cnc and V363~Aur 
within the region allowed by the theoretical constraints for 
stable mass transfer. 

\section*{\sc Acknowledgements}
We thank N. Samus and E. Zhang for providing eclipse timings 
of AC~Cnc, and Homer Giannakis for help in reducing the photometry.
We are indebted to Tom Marsh for the use of his software packages PAMELA and 
MOLLY, and we thank Uli Kolb, Stuart Littlefair and Tariq Shabaz for useful 
discussions. We also thank the referee, Robert Smith, for his careful 
reading of the manuscript and suggestions for improvements. 
TDT and MJS are supported by PPARC studentships; CAW is supported by PPARC 
grant number PPA/G/S/2000/00598. 
DS acknowledges a Smithsonian Astrophysical Observatory Clay Fellowship. 
The INT and JKT are operated on the island of La Palma by the 
Isaac Newton Group in the Spanish Observatorio del Roque de los Muchachos 
of the Instituto de Astrofisica de Canarias.

\bibliographystyle{mnras}
\bibliography{refs}

\end{document}